\documentclass[]{spie}  %>>> use for US letter paper
\usepackage[]{graphicx}

\usepackage{amsmath}
\usepackage{prettyref}
\newrefformat{chap}{Chapter~\ref{#1}}
\newrefformat{sec}{Section~\ref{#1}}
\newrefformat{fig}{Fig.~\ref{#1}}
\newrefformat{eqn}{Eq.~\ref{#1}}
\newrefformat{tab}{Table~\ref{#1}}

\title{On-sky performance during verification and commissioning of 
the Gemini Planet Imager's adaptive optics system} 

\author{Lisa A. Poyneer\supit{a},
Robert J. De Rosa\supit{b,c}, 
Bruce Macintosh\supit{d,a}, 
David W. Palmer\supit{a},
Marshall D. Perrin\supit{e},
Naru Sadakuni\supit{f}, 
Dmitry Savransky\supit{g}
Brian Bauman\supit{a},
Andrew Cardwell\supit{f}, 
Jeffrey K. Chilcote\supit{h},
Daren Dillon\supit{i},
Donald Gavel\supit{i},
Stephen J. Goodsell\supit{f}, 
Markus Hartung\supit{f}, 
Pascale Hibon\supit{f}, 
Fredrik T. Rantakyr\"o\supit{f}, 
Sandrine Thomas\supit{j,k},
Jean-Pierre V{\'e}ran\supit{l}\skiplinehalf
\supit{a}Lawrence Livermore National Lab, 7000 East Ave., Livermore, CA 94551, USA\skiplinehalf
\supit{b}School of Earth and Space Exploration, Arizona State University, 
PO Box 871404, Tempe, AZ 85287, USA\skiplinehalf
\supit{c}School of Physics, College of Engineering, Mathematics and Physical Sciences, 
University of Exeter, Stocker Road, Exeter, EX4 4QL, UK\skiplinehalf
\supit{d}Kavli Institute for Particle Astrophysics and Cosmology, Stanford University, Stanford, CA 94305, USA\skiplinehalf
\supit{e}Space Telescope Science Institute, 3700 San Martin Drive, Baltimore MD 21218 USA\skiplinehalf
\supit{f}Gemini Observatory, Casilla 603, La Serena, Chile\skiplinehalf
\supit{g}Sibley School of Mechanical and Aerospace Engineering, Cornell University, Ithaca, NY 14853, USA \skiplinehalf
\supit{h}Department of Physics and Astronomy, UCLA, Los Angeles, CA 90095, USA\skiplinehalf
\supit{i}University of California Observatories, UC Santa Cruz, 1156 High Street, Santa Cruz, CA 95064, USA\skiplinehalf
\supit{j}NASA Ames Research Center,  Moffett Field, CA 94035, USA\skiplinehalf
\supit{k}UARC, UC Santa Cruz,  1156 High Street, Santa Cruz, CA 95064 USA\skiplinehalf
\supit{l}National Research Council of Canada Herzberg, 5071 West Saanich Road, Victoria, BC V9E 2E7, Canada \\
}

\authorinfo{Send correspondence to Lisa Poyneer: poyneer1@llnl.gov, 1 925 423 3360}
\begin{document} 
  \maketitle 

%%%%%%%%%%%%%%%%%%%%%%%%%%%%%%%%%%%%%%%%%%%%%%%%%%%%%%%%%%%%% 
\begin{abstract}
The Gemini Planet Imager instrument's adaptive optics (AO) subsystem was designed
specifically to facilitate high-contrast imaging. It features several
new technologies, including computationally efficient wavefront reconstruction with
the Fourier transform, modal gain optimization every 8 seconds, and the spatially filtered wavefront sensor.
It also uses a Linear-Quadratic-Gaussian (LQG) controller (aka Kalman filter)  for both
pointing and focus. We present on-sky performance results from verification 
and commissioning runs from December 2013 through May 2014. The efficient reconstruction
and modal gain optimization are working as designed. The LQG controllers effectively notch out vibrations.
The spatial filter can remove aliases, but we typically use it oversized by about 60\% due to stability problems.
\end{abstract}

%>>>> Include a list of keywords after the abstract 


%%%%%%%%%%%%%%%%%%%%%%%%%%%%%%%%%%%%%%%%%%
%%%%%%%%%%%%%%%%%%%%%%%%%%%%%%%%%%%%%%%%%%
%%%%%%%%%%%%%%%%%%%%%%%%%%%%%%%%%%%%%%%%%%
\section{INTRODUCTION} \label{sec:intro}  % \label{} allows reference to this section

The Gemini Planet Imager (GPI)~\cite{Macintosh12052014} is a hyper-spectral 
coronagraphic instrument specifically designed to directly image exoplanets.
GPI and the SPHERE instrument~\cite{doi:10.1117/12.790120} are part of a 
new generation of instruments that feature specialized adaptive optics (AO)
systems that use thousands of actuators and specialized algorithms. In this paper we
discuss the algorithms and technologies that were developed for GPI, and how
well they are actually working during the instrument's  verification and 
commissioning period at Gemini South (late 2013-summer 2014).

%%%%%%%%%%%%%%%%%%%%%%%%%%%%%%%%%%%%%%%%%%
%%%%%%%%%%%%%%%%%%%%%%%%%%%%%%%%%%%%%%%%%%
%%%%%%%%%%%%%%%%%%%%%%%%%%%%%%%%%%%%%%%%%%
\section{Experimental procedures and data analysis} \label{sec:methods}  

The performance of the AO system is analyzed primarily through system diagnostic data.
This telemetry is dumped at the full frame rate of the AO system and can include
anything from raw CCD frames, to calculated centroids, to reconstructed phases to
deformable mirror commands. By synchronizing multiple data dumps, long streams
of telemetry can be saved during on-sky experiments.
During a typical on-sky experiment, 22-second streams of telemetry are saved. The 
specific control parameters in use (e.g.  filter coefficients) are also saved. 
These data are analyzed offline using a standard telemetry analysis code.

Our primary analysis technique is to examine the temporal power spectral densities (PSDs)
of specific modes that the system controls. For tip, tilt and focus we analyze the coefficients
directly. For the higher order phase, we begin with the residual phase as seen by the 
AO system directly after phase reconstruction from the centroids. For a given 22-second interval, the
$x$-$y$-$t$ data cube is converted at each time step into the frequency domain to produce a 
 $f_x$-$f_y$-t data cube. Given the time series of a modal coefficient 
  (e.g. tip, Fourier mode $k=12$, $l=12$, where $k$ and $l$ indicate the index
  into a $48 \times 48$ FFT of the phase), the temporal PSD is estimated by
 the averaged modified periodogram technique~\cite{DSP}.
 
\begin{figure}
   \begin{center}
   \begin{tabular}{c}
  \includegraphics[height=5cm]{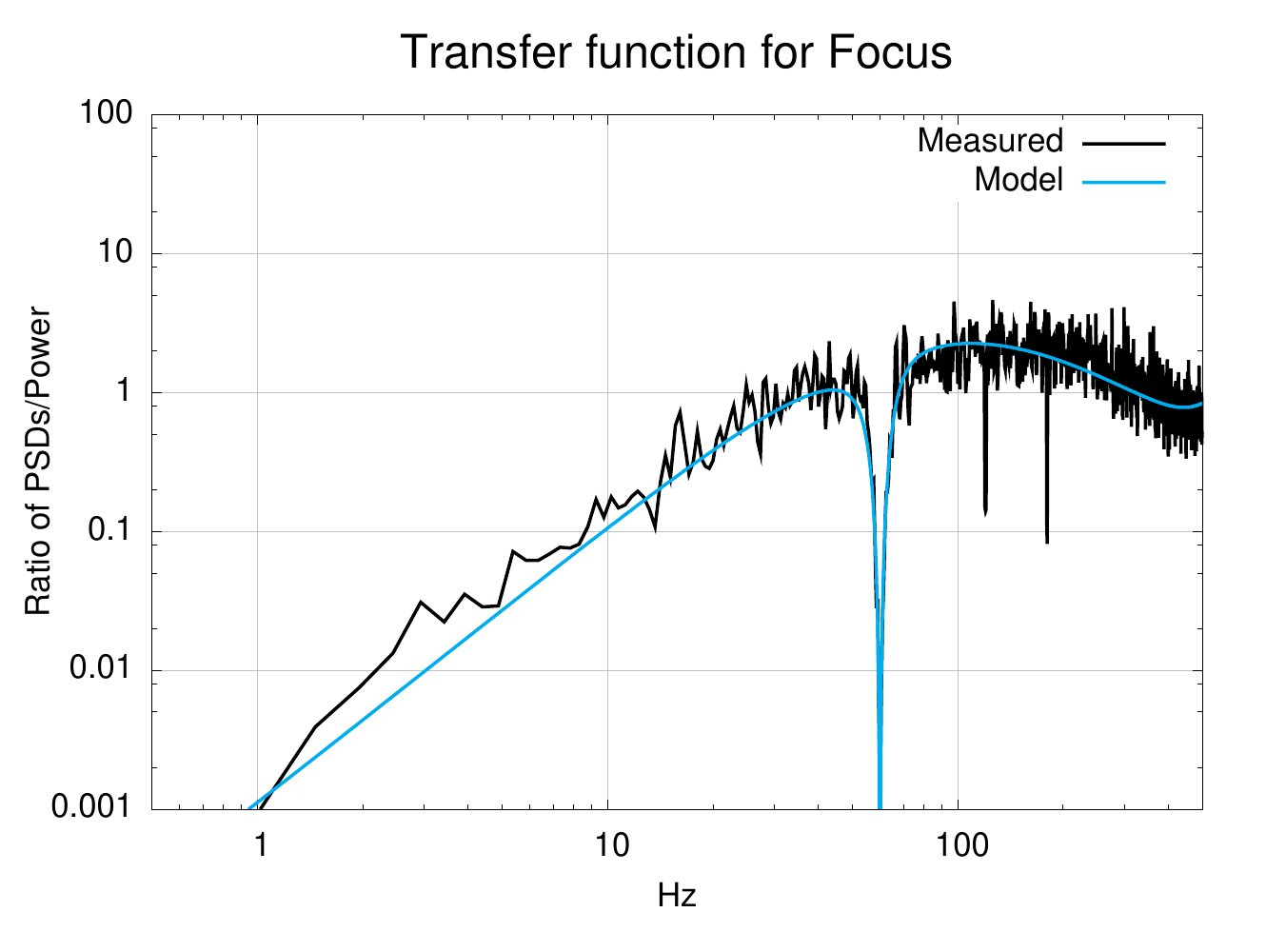}
  \includegraphics[height=5cm]{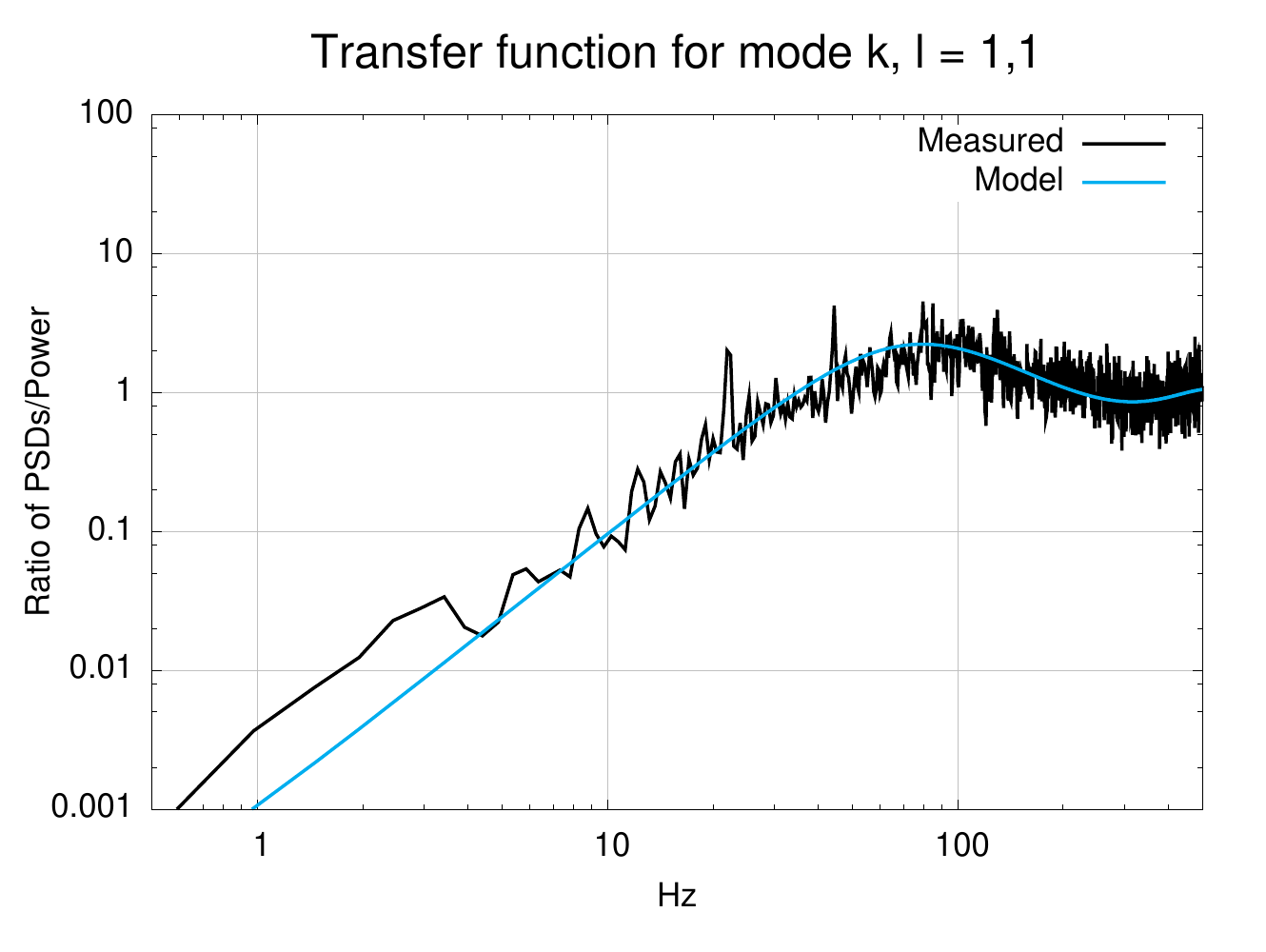}
   \end{tabular}
   \end{center}
   \caption[example] 
%>>>> use \label inside caption to get Fig. number with \ref{}
   { \label{fig:etf} 
The temporal performance of the AO system is well-characterized;
measured error transfer functions using the internal source agree with models. For [left] focus and [right]
a low-order Fourier mode, the measured response on the internal source is given
in black, and the model in blue. Telemetry: \_When\_2014.5.9\_18.54.42}
   \end{figure}

 This temporal PSD represents what is measured by the AO system in closed loop. For bright targets
wavefront sensor (WFS) noise is small and this is very close to the closed-loop error that should be presented 
at the output of the AO system to the science leg. On moderate and dim targets, however,
the measurement and the error are not the same. To analyze the error, as opposed to the measurement, 
we use the exact same technique that the gain optimizer (see below) uses~\cite{Poyneer:OFC}.
To estimate the error, we take the temporal
PSD of the measurement and invert by the known system response. This is possible because the 
temporal response of the system is very well characterized. As shown in Figure~\ref{fig:etf},
the temporal behavior for  focus and a Fourier mode both closely follow our models.
%-------------
%-------------

Once the PSD of the closed-loop measurement is converted into the PSD of the estimated open-loop
measurements, we then split it into signal and noise components. The noise component
should be temporally white, and it is estimated by taking a median of the PSD at the
highest frequencies. This estimate is the PSD of the input noise seen by the AO system. 
The PSD of the signal (primarily atmosphere) seen by the AO system is obtained
by subtracting the noise PSD from the estimated open-loop measurement PSD.
We now have estimated temporal PSDs for the signal that the AO system wants to correct
and the WFS noise that it measures. The error produced by the AO system is estimated
by apply the error transfer function (ETF) to the signal PSD and the noise transfer
function (NTF) to the noise PSD.
This frequency domain method ignores any static level of error, which is removed
before the periodogram calculation. As such (see Section~\ref{sec:eb}) this term is calculated
separately. Note that throughout this paper we rely on the self-reporting of the AO system for
performance analysis. A provisional error budget is given in Section~\ref{sec:eb},
but a rigorous comparison of science measurements to AO telemetry is beyond the scope of this
present work, but will be included in a future publication.

%%%%%%%%%%%%%%%%%%%%%%%%%%%%%%%%%%%%%%%%%%
%%%%%%%%%%%%%%%%%%%%%%%%%%%%%%%%%%%%%%%%%%
%%%%%%%%%%%%%%%%%%%%%%%%%%%%%%%%%%%%%%%%%%
\section{Fourier transform reconstruction} \label{sec:ftr}  % \label{} allows reference to this section

%%-----------------------------------------------------------
%%-----------------------------------------------------------
\subsection{Technology description and methods} 

The concept of reconstructing the wavefront phase from slopes with 
a filtering approach was originally proposed in the mid 1980s~\cite{Freisch1,Freisch2}.
However, the boundary problem imposed by a circular (or annular) aperture in the 
square computational grid prevented the method from practical use. By solving
this boundary problem, we were able to make the filtering approach feasible, calling
the method Fourier Transform Reconstruction (FTR)~\cite{Poyneer:FTR_JOSA}.
We further defined new filters that better captured Shack-Hartmann behavior~\cite{Poyneer:FTR_SPIE02,Poyneer:OFC} and showed that the Fourier basis 
provided a tight frame for modal control~\cite{Poyneer:OFC}. When formulated to consider
the signal and noise power levels, the FTR method is a Wiener filter~\cite{Poyneer2007Signal-processi}, 
and we showed that
FTR is equivalent~\cite{Poyneer2007Signal-processi} to the Minimum Variance 
Unbiased approach~\cite{MVU} if the Fourier basis  set is used.
To our knowledge, the use of FTR in GPI represents the first on-sky use of a 
computationally efficient wavefront reconstruction method (as opposed to 
a parallelized vector-matrix-multiply) in an astronomical AO system.

After phase reconstruction the residual is split~\cite{Lavinge2007Woofer-Tweeter-} 
between two deformable mirrors (DMs).
Low-spatial-frequency Fourier modes, which require the most stroke,  
are sent to the ``Woofer'' DM, a conventional
piezo DM manufactured by Cilas. The high-order phase is sent to the 
``Tweeter'' DM, a microelectromechanical systems (MEMS) mirror developed
by Boston Micromachines.

%%-----------------------------------------------------------
%%-----------------------------------------------------------
\subsection{On-sky results} 

Use of FTR was motivated by computational complexity; a matrix multiplication of
a system of the same size would require 45 times more computation. This enabled
the implementation of the real-time controller (RTC) on a commercial, off-the-shelf (COTS)
 server running real-time
linux, as opposed to customized or special-purpose hardware such as digital signal processors (DSPs)  or
graphics processing units (GPUs).
As we see below, the use of FTR makes the residual Fourier coefficients
of the wavefront available, which facilitates computationally efficient controller
optimization. As of May 2014, the total delay from the end of a WFS frame integration
to the application of the phase commands on the DMs is 1.7 msec. This is true for either
1 kHz or 500 Hz operation. The tip, tilt and focus commands are computed more
quickly since they do not go through the FTR process. Those commands are placed on the 
Woofer DM and the tip-tilt stage after 1.2 msec. 
For both of these with first 890 microsec are taken up by the detector read.

As has been noted by us elsewhere\cite{Poyneer2006Wavefront-contr}, 
use of FTR requires precise alignment of the AO system optics.
This alignment is achieved through automatic processes before each
acquisition~\cite{Savransky:13}. The rotation of the spots grid produced by 
the lenslets relative to the WFS CCD pixel grid has proved problematic. 
As was originally noted in Section 5.2 of our prior work~\cite{Poyneer2006Wavefront-contr},
if rotation exists in the centroids, it will be reconstructed by FTR into a 
shape that has a scalloping pattern around the edges. During closed-loop
operation this behaves similarly to an alias - a non-physical signal is
present in the slopes and makes it through to the DM, where a shape
builds up to attempt to correct it.

As built,  there is a small but not insignificant amount of rotation in the WFS, which is nominally 
included in the reference slopes. The removal of rotation via reference subtraction
will not work completely if some effect changes the gain of the 
quad cells. This could be a drift of the bias level on the CCD~\cite{sadakunithis}, 
spot size variations due to atmospheric turbulence, or
potentially other subtle factors. 
During integration and testing we identified that bias drift was significantly reducing
performance quality; the short-term solution was taking regular dark frames.
Despite regular darks, during the December run we identified that rotation was not 
being fully removed on-sky, most likely due to spot size variations that are
dependent on seeing. These spot size variations result in a large status excursion of the 
Tweeter actuators from flat around the edges of the pupil. 
To solve this problem the RTC now includes a step where rotation is explicitly 
removed from the centroids before FTR. This approach has been verified to 
work correctly, and has reduced the edge excursions. Pending results from
other analysis by Greenbaum~\cite{greenbaumthis} and Sadakuni~\cite{sadakunithis} 
the specific vectors used in the removal process may be modified.

This is our best solution to date for this problem with FTR. On a symmetric pupil
this effect is entirely due to the edge correction process where the slopes
on the square grid are managed. An equivalent matrix reconstructor on a 
symmetric pupil would not let any rotation through. As a further complication,
GPI does not have a symmetric pupil; small subregions are masked off to 
prevent the AO system from seeing the dead actuators, which are mostly positioned
at the exterior edge of the pupil. This asymmetry allows for rotation to leak
through both FTR and a matrix reconstructor. If we were using a matrix method,
we would not need to do the removal step from the centroids but could instead
account for the rotation internally in the matrix.

%%%%%%%%%%%%%%%%%%%%%%%%%%%%%%%%%%%%%%%%%%
%%%%%%%%%%%%%%%%%%%%%%%%%%%%%%%%%%%%%%%%%%
%%%%%%%%%%%%%%%%%%%%%%%%%%%%%%%%%%%%%%%%%%
\section{Modal gain optimization} \label{sec:ofc}  % \label{} allows reference to this section

%%-----------------------------------------------------------
%%-----------------------------------------------------------
\subsection{Technology description and methods}

Originally proposed by Gendron and Lena~\cite{GEND-94}, modal gain optimization 
is a technique where the wavefront phase is decomposed into orthogonal modes (as opposed to 
actuator positions in the pupil). Using either open- or closed-loop telemetry 
on-sky, the control loop gain of each mode is set such that it minimizes the 
residual error variance. Modal gain optimization has 
been implemented in several AO systems, including the initial demonstration in
Come-On-Plus~\cite{GEND-95}, which used open-loop telemetry prior to an observation.
The method is used in the Altair AO system at Gemini North~\cite{AltairOpt} and 
ESO's NAOS system~\cite{ROUS-03}. More recently, modal gain optimization 
for up to 400 modes has been implemented in the LBT's FLAO system~\cite{doi:10.1117/12.927109}.
As described, that optimization is done once at the beginning of an observation, and proceeds iteratively.

In GPI, the modal gain optimization procedure operates continuously as 
a supervisory process. As described fully elsewhere~\cite{Poyneer:OFC}, the 
Fourier coefficients of the residual phase are buffered during the reconstruction
process. Each mode is analyzed temporally, using the averaged-modified periodogram
technique to estimate the closed-loop power spectral density (PSD) of the measurements.
Using a fast root-finding technique, the best modal gain for each Fourier mode is calculated 
independently. All 2304 modal gains (which have hermitian symmetry on the 
spatial frequency grid, so only half need to be calculated) 
are used in a gain filter which is applied during the reconstruction
process. 
Because the Fourier modal coefficients are directly available in closed loop, and because
changing the modal gains involves simply changing a 2304-element vector in memory,
the Fourier framework of GPI allows modal gain optimization to be conducted every
eight seconds for all 2304 Fourier modes during closed loop. As such, we consider GPI to be
 unique in terms of the large number of modes that it optimizes, and the fact that it does so
continuously during science observations.

%%-----------------------------------------------------------
%%-----------------------------------------------------------
\subsection{On-sky results}

\begin{figure}
   \begin{center}
   \begin{tabular}{c}
  \includegraphics[height=4cm]{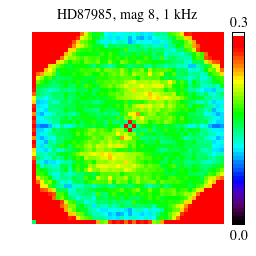}
  \includegraphics[height=4cm]{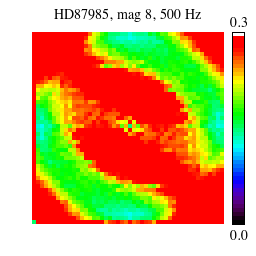}
  \includegraphics[height=4cm]{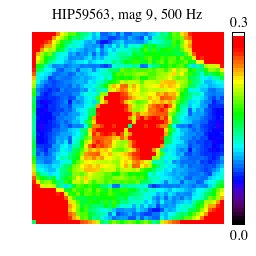}
  \includegraphics[height=4cm]{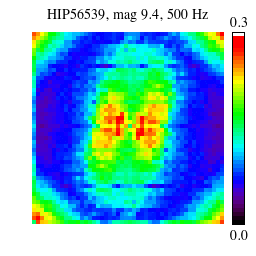}
   \end{tabular}
   \end{center}
   \caption[example] 
%>>>> use \label inside caption to get Fig. number with \ref{}
   { \label{fig:ofc_gains} 
Modal gain filters as determined during closed-loop operation on sky. Each
panel color scale goes from 0.0 to 0.3 for the gain. Piston ($k=0$, $l=0$) is
at middle; waffle ($k=24$, $l=24$) is at the corner. Wind direction is 
visible, particular in the two cases at right.
Telemetry sets: \_When\_2014.3.24\_1.56.52, \_When\_2014.3.24\_1.36.1, 
\_When\_2014.3.24\_2.37.15 and \_When\_2014.3.24\_3.3.31.}
   \end{figure}

During the integration and testing phase, the modal gain optimization was
verified through off-line re-analysis of telemetry. That is, our analysis codes estimated the 
modal gains from telemetry, and the results were compared to what the 
system itself had calculated.
The primary concern of our on-sky testing is whether the modal gain optimizer
improves overall AO performance. To test this, we conducted a series of experiments.
For a given star magnitude and frame rate, we closed the AO loops with a 
fixed value of uniform modal gain. After taking telemetry 
we turn on the gain optimizer and let the gains settle, a process that 
takes less than 30 seconds. We then take another telemetry set.
We repeat this uniform gain-optimized gain pairing several times, slowly stepping
up the initial uniform gain from 0.05 to 0.3 with increments of 0.05. This gives us an
interleaved set of measurements over a period of about 20 minutes, with 
six examples of uniform gain filters spanning the range of stable gains and 
six examples of optimized gains. During our March run we conducted
four of these tests. Representative gain filters for the four cases are shown in 
Figure~\ref{fig:ofc_gains}. As predicted by simulations~\cite{Poyneer2006Optimal-Fourier}, the gain filters span 
a wide range of possible gains, and show evidence of wind direction.
%-------------
%-------------

\begin{figure}
   \begin{center}
   \begin{tabular}{c}
  \includegraphics[height=5cm]{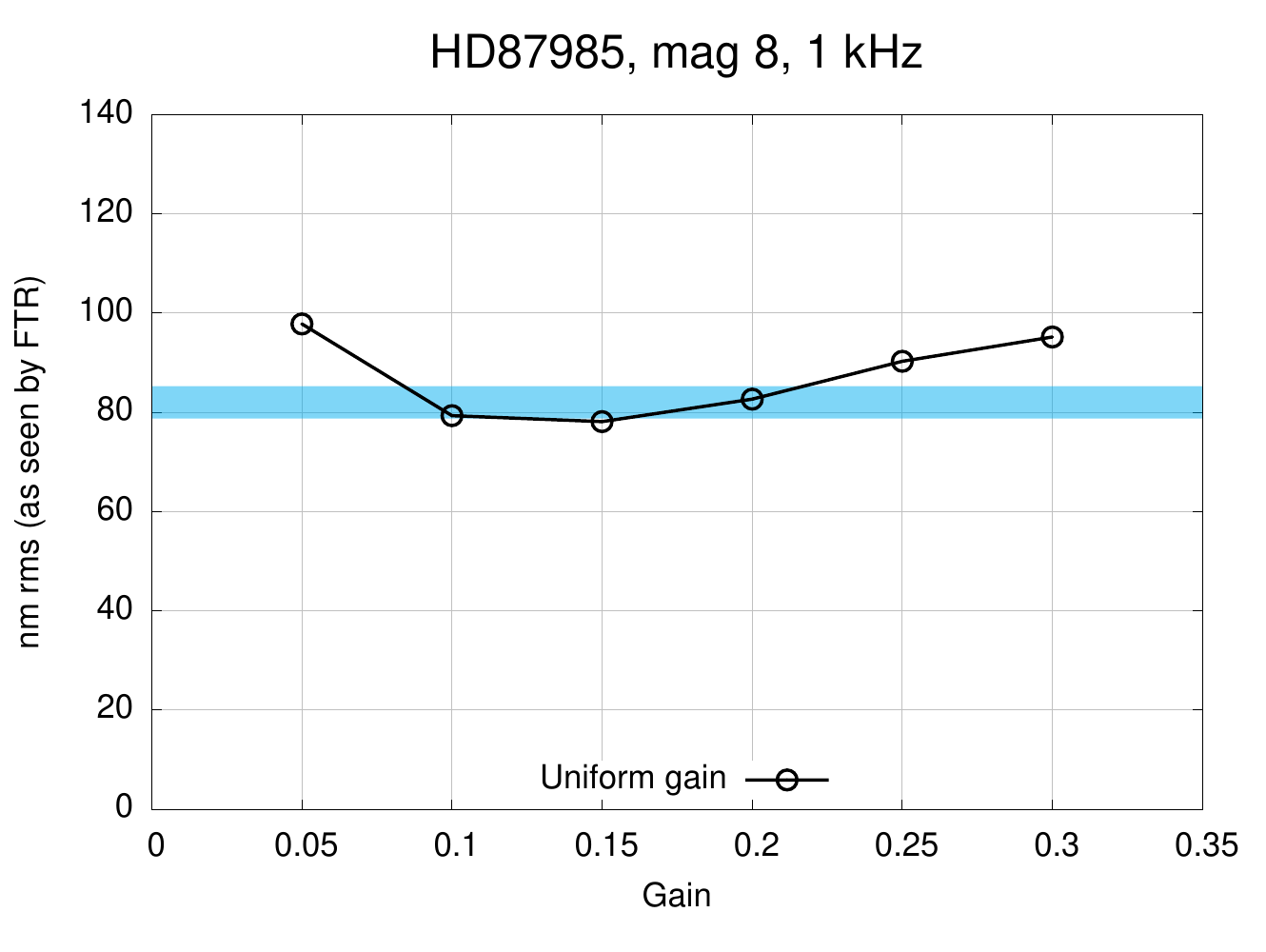}
  \includegraphics[height=5cm]{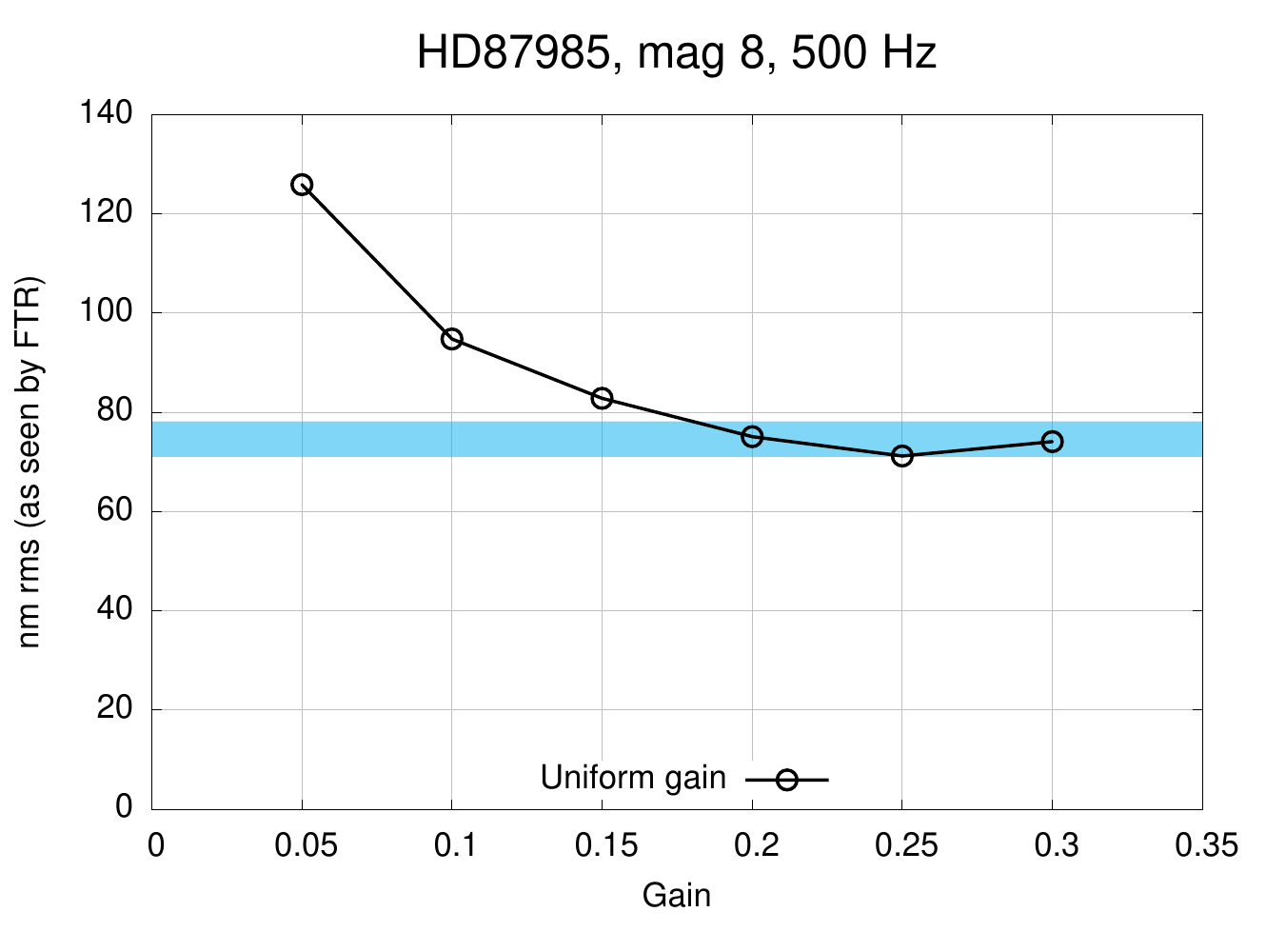}\\
  \includegraphics[height=5cm]{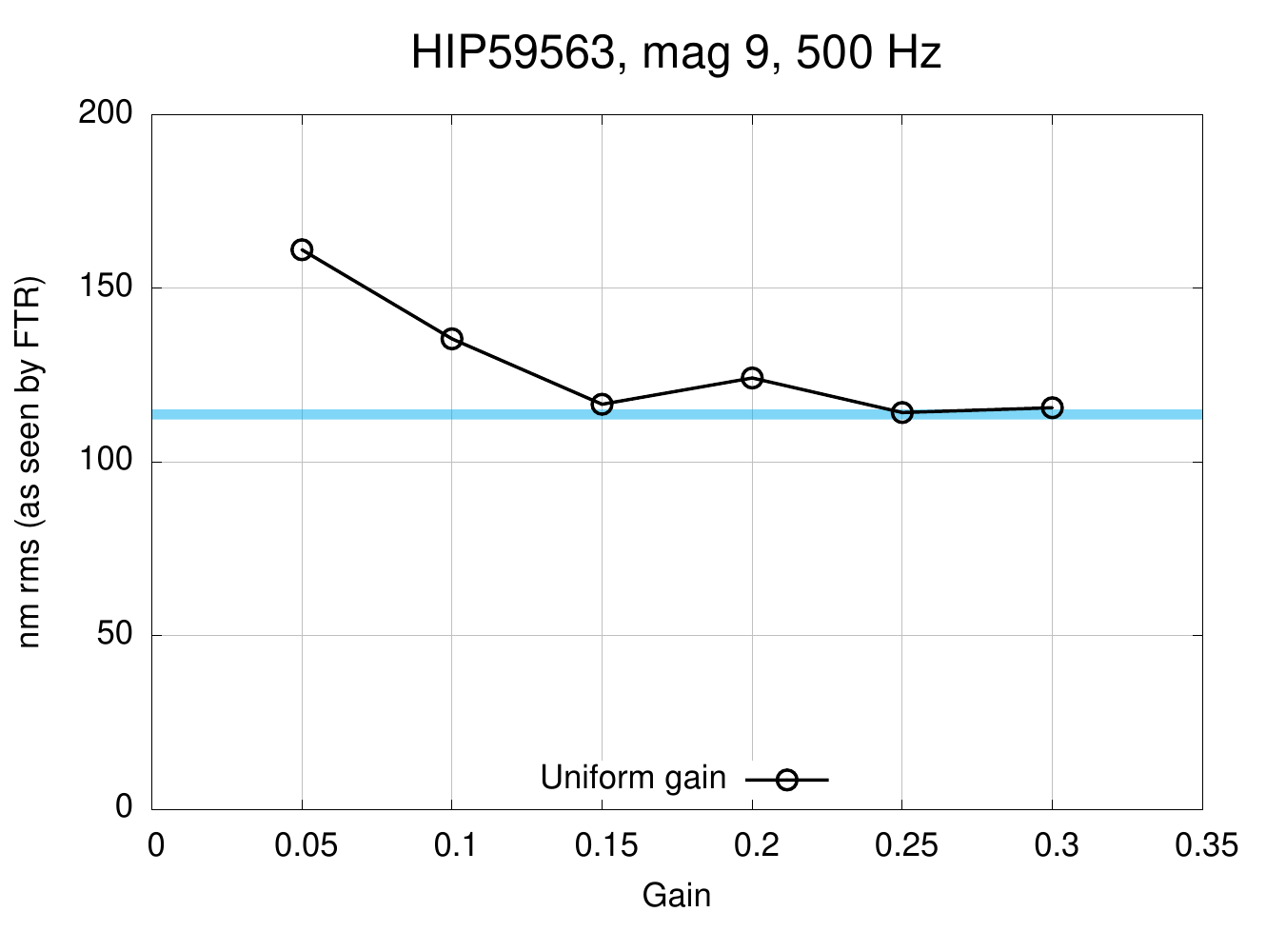}
  \includegraphics[height=5cm]{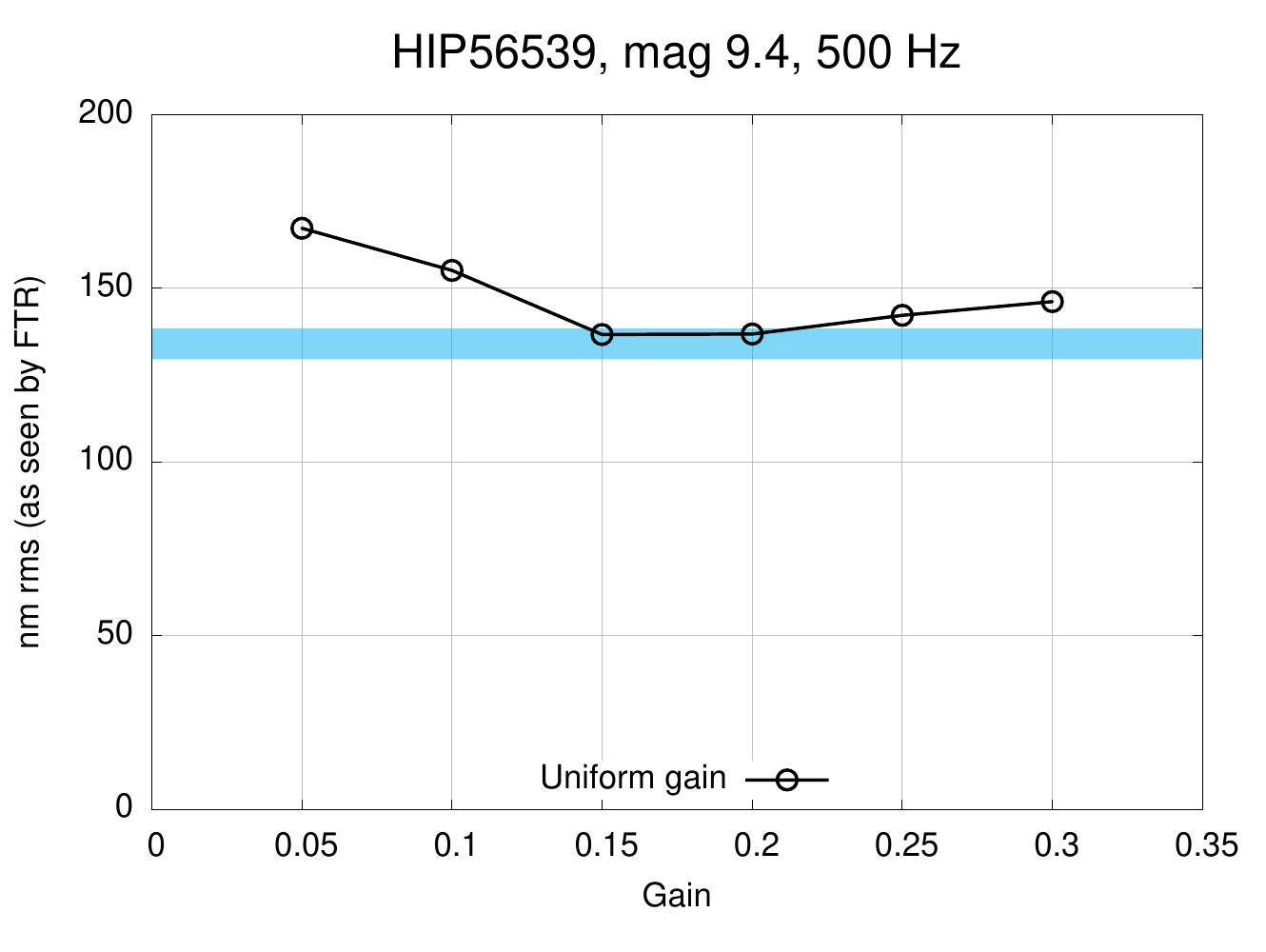}
   \end{tabular}
   \end{center}
   \caption[example] 
%>>>> use \label inside caption to get Fig. number with \ref{}
   { \label{fig:ofc_overall_4targets} 
Error estimated from AO measurements, for three different targets for the optimizer test.
In each case black points represent the six uniform gain trials; blue band represents full range of
error for the six optimizer trials. Best performance on target HD87985 comes 
with optimization at 500 Hz, not at 1 kHz. Telemetry: 48 different telemetry sets used spanning 
the period \_When\_2014.3.24\_1.25.48 to \_When\_2014.3.24\_3.13.12.}
   \end{figure}

Using the methods described  in Section~\ref{sec:methods}, we estimate the 
error for each Fourier mode and for all modes together.
Figure~\ref{fig:ofc_overall_4targets} 
%-------------
%-------------
shows the estimates of total error. For each of the six uniform gain cases, the total error is
marked by the black points. This total error is the sum of the error due to signal (which
decreases as gain is increased) and the error due to noise (which increases as gain is increased).
Analysis of the two components (not shown) confirms that the system follows
this expected behavior. In the figure the blue bar indicates the full range of the 
total error, as estimated from the six different optimized cases. In all these cases the 
performance with the optimizer on (in blue) is at or very close to the minimum
when the gain is adjusted by hand.

\begin{figure}
   \begin{center}
   \begin{tabular}{c}
  \includegraphics[height=5cm]{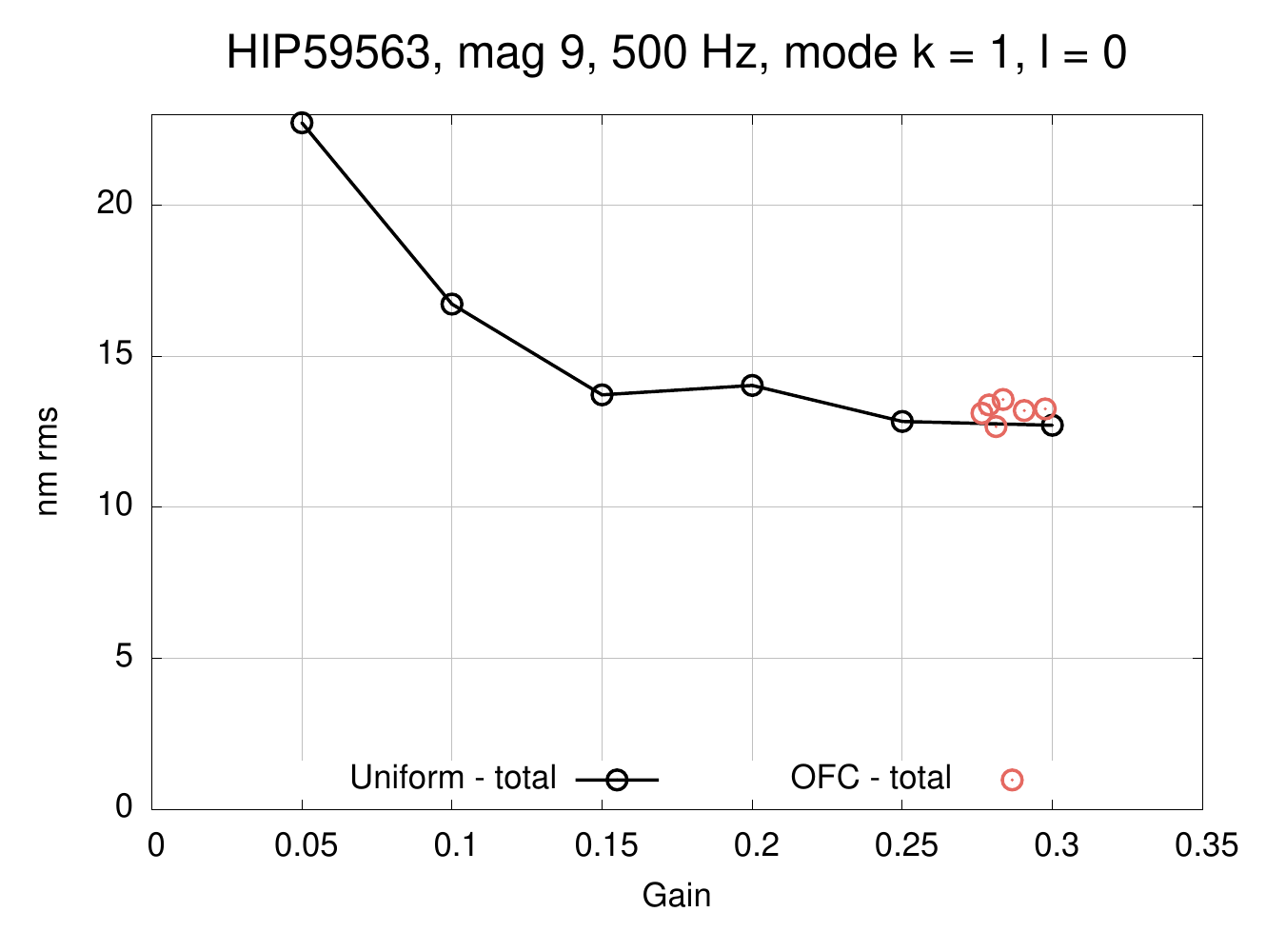}
  \includegraphics[height=5cm]{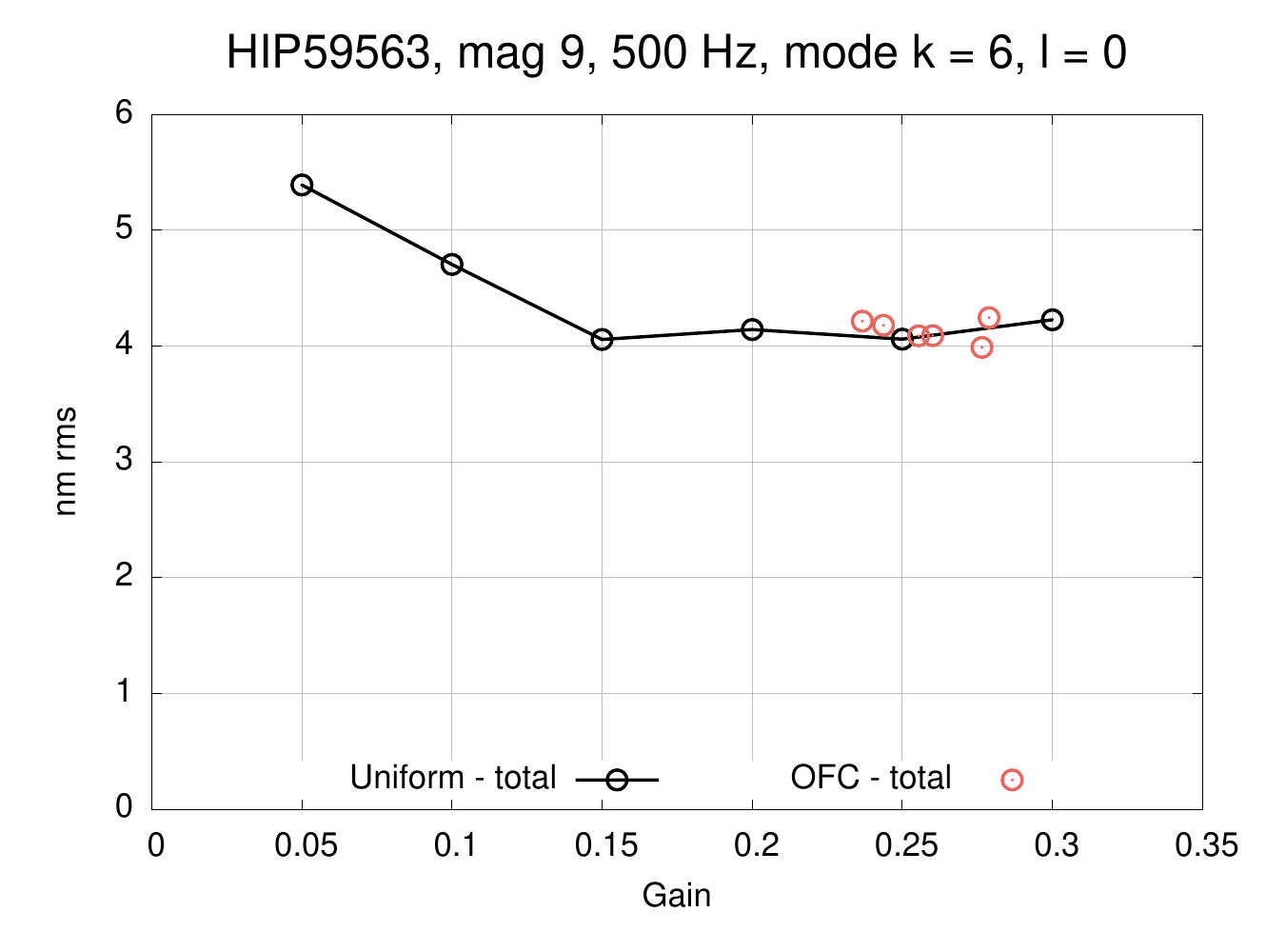}\\
  \includegraphics[height=5cm]{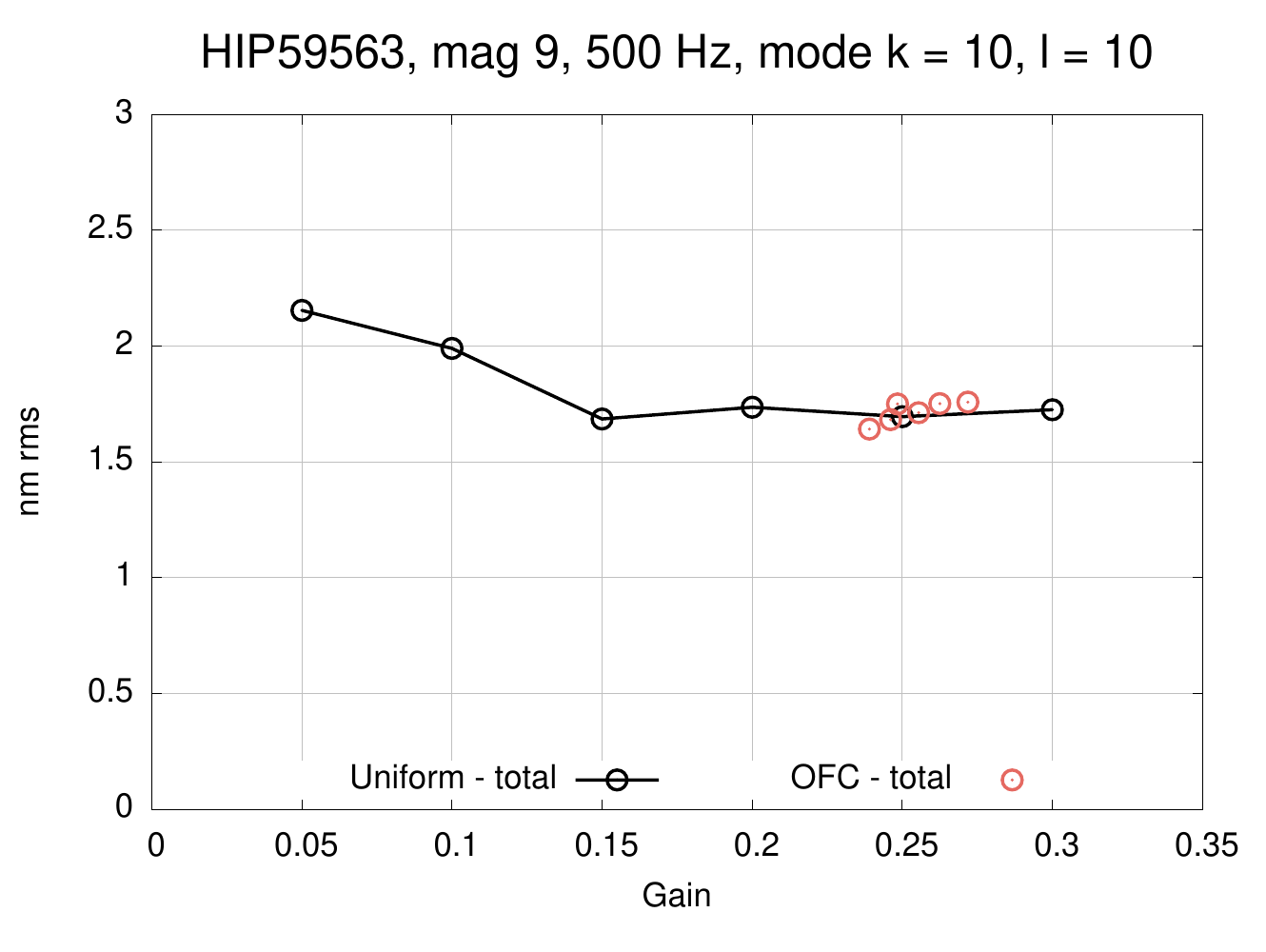}
  \includegraphics[height=5cm]{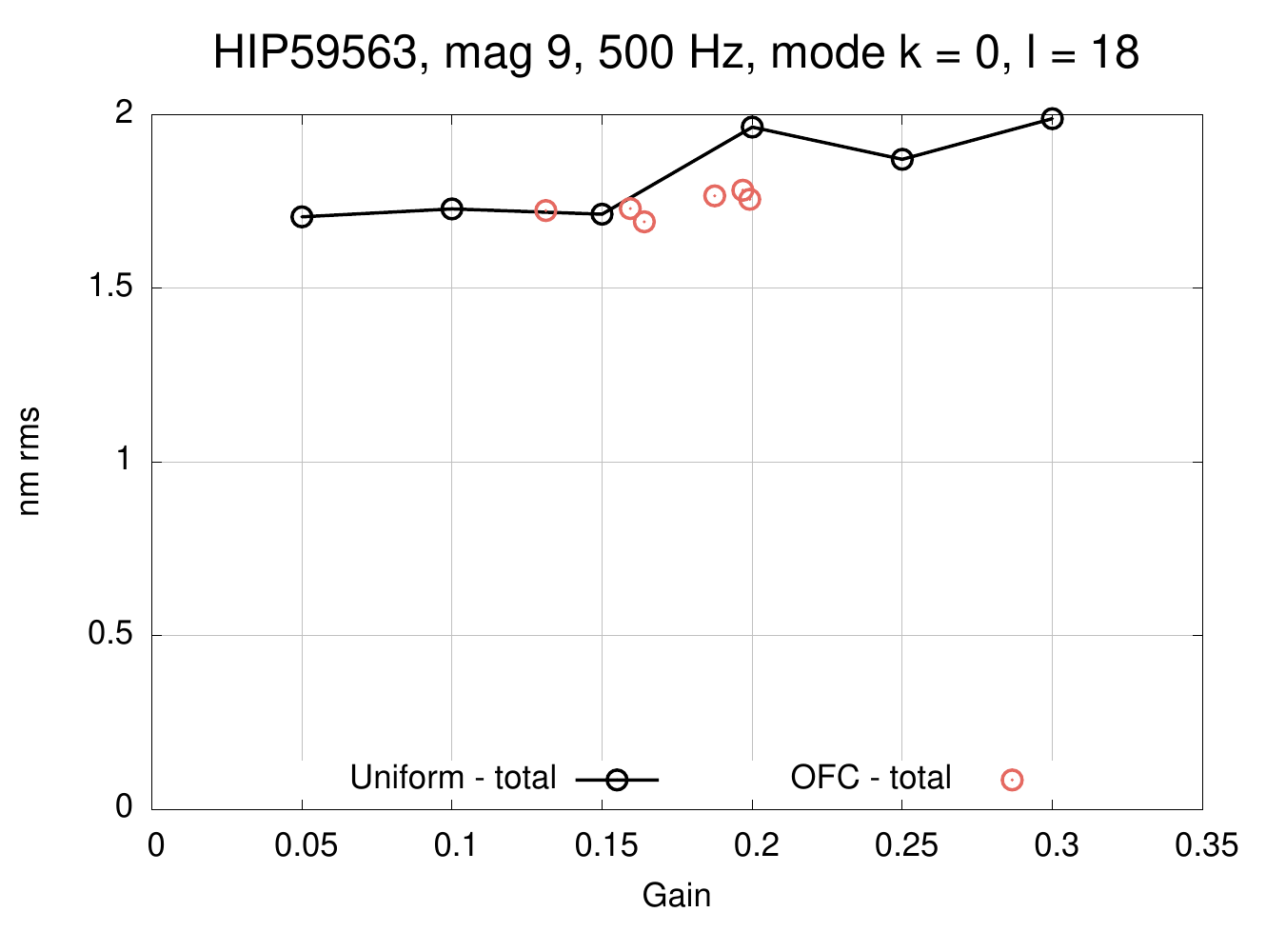}
   \end{tabular}
   \end{center}
   \caption[example] 
%>>>> use \label inside caption to get Fig. number with \ref{}
   { \label{fig:ofc_modes} 
Estimated science-leg error from AO residuals, for four different Fourier modes (mode
index given in plot titles). Black points are for the six uniform gain trials; red points are for the 
six optimized gain trials. Low-order modes (upper left) have a large amount of power, while 
higher-frequency modes have less. On a mode-by-mode basis the gain optimizer is correctly
finding gains which minimize error. Telemetry: 12 different sets used spanning 
the period \_When\_2014.3.24\_2.27.11 to \_When\_2014.3.24\_2.47.5}
   \end{figure}

This total error metric shown in Figure~\ref{fig:ofc_overall_4targets} 
 is dominated by the optimizer's performance at
low spatial frequencies. This is because both 
atmospheric turbulence and the noise propagation
from the Shack-Hartmann wavefront sensor produce more power at lower
frequencies.  To further verify that
the modal gain optimization is working for all modes, we analyzed each mode individually. 
Figure~\ref{fig:ofc_modes} shows the test results for four different Fourier modes
that are representative of the 2304 controllable modes.  
%-------------
%-------------
As shown in this figure, for a 9th magnitude star at 500 Hz, the lowest spatial frequency
modes (e.g. $k=1$, $l=0$ shown at upper left) have the most total error. 
In this case the atmosphere dominates and the optimizer correctly drives the 
modal gain to maximum. For mid-frequency modes, such as $k=6$, $l=0$  at upper right
and $k=10$, $l=10$  at lower left, the atmosphere and noise are more
equal. For these modes the optimizer uses a high, but not maximum gain. 
For the highest spatial frequencies, such as $k=0$, $l=18$ shown at bottom right,
noise dominates and the optimizer drives the gain lower. These modes also have significant 
high-temporal frequency content due to aliasing (see below), which the optimizer
does not explicitly know about.

In order to suppress waffle and reduce edge effects, during Integration and Test
a modified Tweeter influence function filter was put in place. As discussed 
elsewhere~\cite{Poyneer2011The-use-of-a-hi}, 
the estimated phase is pre-compensated by 
a influence function filter to correctly shape the phase on the Tweeter. 
Due to the broadness of the Tweeter influence function, the
effective Tweeter gain for high spatial frequencies~\cite{morzinski:627221}
is very small. Division
by this small number inflates noise. As such, we artificially limit the division. As a result,
the optimizer tends to drive up the gain on these under-compensated modes, as is 
obvious by the red regions in the corners of the gain filters shown in Figure~\ref{fig:ofc_gains}.
This behavior of the optimizer to correct for a miscalibration was discussed
in Section 4.C of our original proposal~\cite{Poyneer:OFC}. It remains to 
be determined if artificially suppressing the influence function gains
and then having the optimizer try to crank them back up  is adversely affecting performance on-sky.

%\clearpage

%%%%%%%%%%%%%%%%%%%%%%%%%%%%%%%%%%%%%%%%%%
%%%%%%%%%%%%%%%%%%%%%%%%%%%%%%%%%%%%%%%%%%
%%%%%%%%%%%%%%%%%%%%%%%%%%%%%%%%%%%%%%%%%%
\section{Spatially-filtered wavefront sensor} \label{sec:sfwfs}  % \label{} allows reference to this section

%%-----------------------------------------------------------
%%-----------------------------------------------------------
\subsection{Technology description and methods} 

Aliasing occurs when a signal has frequency content above half
of its sampling frequency. In an adaptive optics system, the atmosphere
is not-band-limited; this will cause any sensor which samples the phase,
either through lenslets or pixels, to measure spurious signals.
These aliases will lead to extra residual error. Rigaut first calculated this
as one-third of the classical fitting error~\cite{RigautModel}.
PSD treatments such as those by Jolissaint~\cite{Jolissaint2006Analytical-mode}
show that the exact amount of aliasing error depends on the filter type
(derivative vs. direct phase) and the control parameters.

To prevent the aliasing error, we has proposed what we termed the 
spatially-filtered wavefront sensor (SFWFS)~\cite{Poyneer:SFWFS_JOSA}.
 For a Shack-Hartmann sensor,
this is implemented as a hard-edged field stop in a focal plane of the WFS
before the lenslet array. In the case of a high-strehl system, the filter will
reject phase errors that scatter light beyond the size of the field stop.
In the case of a system with subaperture size $d$, the field stop with 
a diameter of $\lambda/d$ will reject content beyond the sampling 
limit, in theory producing a band-limited phase for the WFS to measure
and producing a dark hole in the PSF.
A first experimental demonstration of the SFWFS was done by Fusco~\cite{Fusco2005Closed-loop-exp},
where as over-sized stop was shown to improve performance in 
a testbed with dynamic turbulence. 
In our own work we were able to demonstrate cleaning out a dark hole 
on static phase plate with a 32x32 MEMS mirror and sensor~\cite{Poyneer2005Experimental-de}.

%%-----------------------------------------------------------
%%-----------------------------------------------------------
\subsection{Modifications from original design} 

During the integration and testing phase of GPI, we had to make some
accommodations in using the SFWFS. In particular we discovered that 
when sized to nominal (i.e. $\lambda/d$), we observed non-linear effects
in edge sub-apertures and in the subapertures near the Tweeter's dead actuators.
Fourier optics simulations revealed that for large phase errors such as 
an actuator stuck more than a micron from its neighbors, the spatial filter
is non-linear and will produce biased measurements. To mitigate this,
we implemented a stronger leak (for an integral controller of the form $g/(1-c\mathrm{z}^{-1})$, $c$ would be set to 0.9) on the actuators immediately next to 
the dead actuators to bleed off this error. Another problem we encountered
is poor stability at the edge of the pupil, which leads to loss of light. During
integration and testing we examined this, and found it was exacerbated by a 
still poorly-understood effect on the partially illuminated subapertures along the 
outer edge of the pupil. During this phase we had also used a stronger leak  ($c = 0.9$)
on the actuators bordering these partially-illuminated subapertures. During
our May 2014 testing run we switched to simply not using measurements from
these partially illuminated subapertures around the outside edge. Operation
remains good in this slightly modified configuration, though it remains
to be determined if this has improved spatial filter stability at all.

%%-----------------------------------------------------------
%%-----------------------------------------------------------
\subsection{On-sky results} 

\begin{figure}
   \begin{center}
   \begin{tabular}{c}
  \includegraphics[height=5cm]{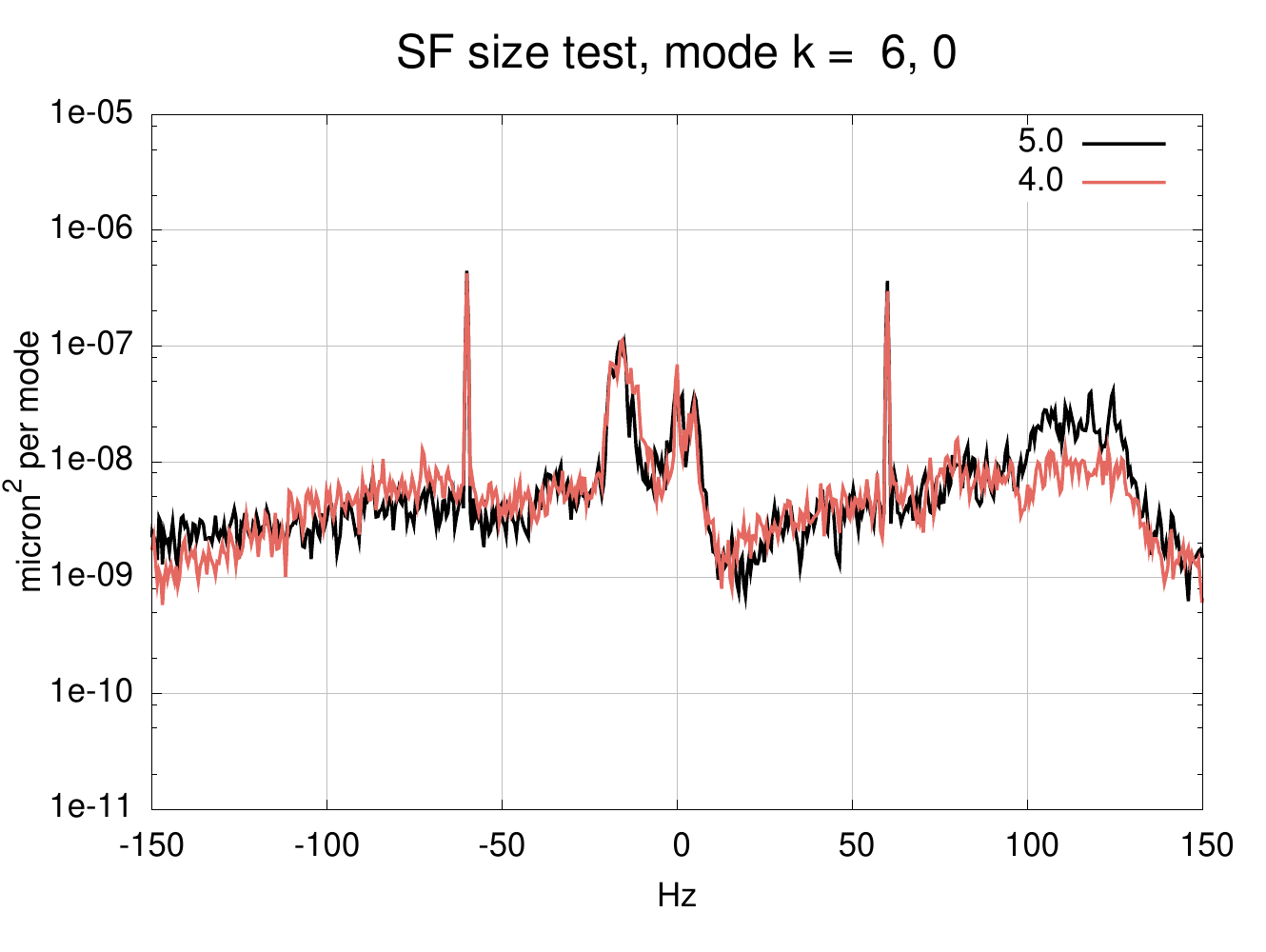}
  \includegraphics[height=5cm]{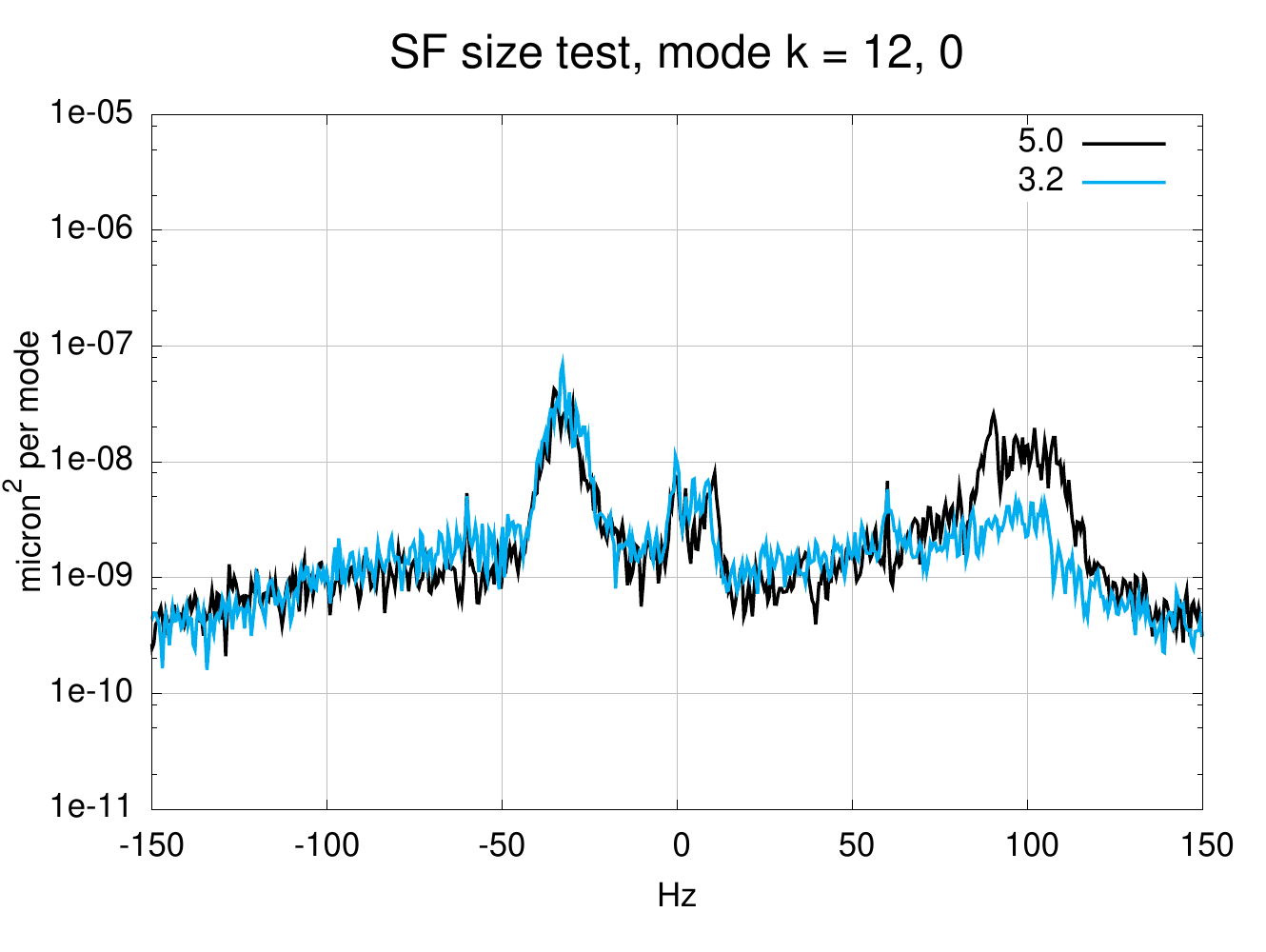}
   \end{tabular}
   \end{center}
   \caption[example] 
%>>>> use \label inside caption to get Fig. number with \ref{}
   { \label{fig:sf_psds} 
Bright star spatial filter test shows that the filter removes aliases of wind-blown turbulence.
At left, the filter at 4.0 mm cleans up modes out to spatial frequency radius 6. At right,
the filter at 3.2 mm cleans up modes out to spatial frequency radius 12. See text for a full explanation.
Telemetry sets: \_When\_2014.5.14\_19.53.43, \_When\_2014.5.14\_19.58.46 and 
\_When\_2014.5.14\_20.3.41 taken on target HD101615.}
   \end{figure}

We can assess the ability of the SFWFS to reject specific spatial frequencies
by examining the temporal PSDs of the Fourier modes during closed loop.
When frozen flow is present (which it typically is), the wind motion creates 
characteristic peaks in the temporal PSDs of the Fourier modes. For any Fourier
mode of spatial frequency (inverse-meters) $<f_x, f_y>$ a wind layer with velocity 
vector (m/s) $<v_x, v_y>$, a peak appears in the PSD at temporal frequency (Hz) 
$f_x v_x + f_y v_y$. For the controllable modes, this creates a characteristic
pattern (see Figure 4 of Poyneer et al.~\cite{Poyneer2008Experimental-ve}). When a 
spatial frequency above the sampling limit of the AO system is measured,
it aliases down to a lower spatial frequency, but the temporal frequency of the 
wind component stays the same. This results in peaks in the controllable modes
from aliasing at temporal frequencies $(f_x \pm d^{-1})v_x + (f_y \pm d^{-1})v_y$,
where $d$ is the subaperture size in the pupil, e.g. 18 cm for GPI. An example of
this is shown in Figure~\ref{fig:sf_psds}, left side. 
%-------------
%-------------
At left the plot shows that temporal PSD of
the residual phase for Fourier mode $k = 6$, $l = 0$. We know, from examining the 
other Fourier modes, that the peak at -20 Hz is from frozen flow atmosphere.
The alias is visible at right from 100 to 130 Hz. In black is the residual 
as seen with the SF = 5.0 mm. When the filter is closed to 4.0 mm, this
alias is suppressed. (The leakage of the non-focus 60 Hz vibrations are clearly seen at 60 Hz
for this Fourier mode). In the right panel we see Fourier mode $k = 12$, $l = 0$.
In this case the wind peak has moved out to temporal frequency -30 Hz. The
alias has moved down to about 90 to 110 Hz. For this higher frequency the blue curve
shows the residual as seen with the SF at 3.2 mm. As the filter gets smaller, it cleans
up aliases for controllable Fourier modes of higher frequency. 

The figure shows data taken on May 14, 2014 on target HD101615. We also have 
data from December 11, 2013 on $\beta$ Pictoris down to SF size 2.8 mm. 
A qualitative analysis of the rejection of aliases of wind-blown turbulence produces estimates for
the range of modes that the SFWFS ``cleans up'' by removing aliases. As given in Table~\ref{tab:sf},
this follows the expected linear trend as a function of SF size.
At size 3.2 mm, the filter cleans up modes to $k=12$, which is halfway out 
to the edge of the dark hole. Though as the filter narrows beyond this and more
modes are cleared, overall performance is reduced due to edge effects that
compromise stability. The $\beta$ Pictoris testing occurred during very good seeing
(see first $\beta$ Pictoris column of ~\ref{tab:eb}). In this case we were able to 
stably close
the filter to 2.8 mm, which we estimate clears out 66\% of the dark hole. This
performance is atypical.
\begin{table}[h]
\caption{\label{tab:sf}Estimates of spatial filter effectiveness for [left] 2013-12-11 on $\beta$ Pictoris
and [right] 2014-5-14 on HD101615. The maximum spatial frequency for which the filter removes aliases is given;
maximum possible value for $k$ is 24.} 
\label{tab:fonts}
\begin{center}       
\begin{tabular}{|l|l|} %% this creates two columns
%% |l|l| to left justify each column entry
%% |c|c| to center each column entry
%% use of \rule[]{}{} below opens up each row
\hline
\rule[-1ex]{0pt}{3.5ex}  SF size (mm) & Spat. freq. \\
\hline
\rule[-1ex]{0pt}{3.5ex}  3.5 & $k = 10$  \\
\hline
\rule[-1ex]{0pt}{3.5ex}  3.3 & $k = 11$  \\
\hline
\rule[-1ex]{0pt}{3.5ex}  3.0 & $k = 14$  \\
\hline
\rule[-1ex]{0pt}{3.5ex}  2.8 & $k = 16$  \\
\hline
\end{tabular}
\begin{tabular}{|l|l|} %% this creates two columns
%% |l|l| to left justify each column entry
%% |c|c| to center each column entry
%% use of \rule[]{}{} below opens up each row
\hline
\rule[-1ex]{0pt}{3.5ex}  SF size (mm) & Spat. freq. \\
\hline
\rule[-1ex]{0pt}{3.5ex}  4.5 & $k = 0$  \\
\hline
\rule[-1ex]{0pt}{3.5ex}  4.0 & $k = 6$  \\
\hline
\rule[-1ex]{0pt}{3.5ex}  3.5 & $k = 10$  \\
\hline
\rule[-1ex]{0pt}{3.5ex}  3.2 & $k = 12$  \\
\hline
\end{tabular}
\end{center}
\end{table} 

%\clearpage

%%%%%%%%%%%%%%%%%%%%%%%%%%%%%%%%%%%%%%%%%%
%%%%%%%%%%%%%%%%%%%%%%%%%%%%%%%%%%%%%%%%%%
%%%%%%%%%%%%%%%%%%%%%%%%%%%%%%%%%%%%%%%%%%
\section{Tip-tilt vibration correction with LQG} \label{sec:ttlqg}  % \label{} allows reference to this section

%%-----------------------------------------------------------
%%-----------------------------------------------------------
\subsection{Technology description and methods} 
Mathematical treatments of optimal control methods for AO 
date back more than a decade, such as Gavel \& Wiberg's~\cite{StrehlOpt}
Kalman filter. Le Roux developed a Linear-Quadratic-Gaussian (LQG) 
framework for AO control\cite{Roux2004Optimal-control}, which was
experimentally demonstrated in a testbed for vibration filtering by Petit~\cite{Petit2008First-laborator}.
Further progress has recently been made with an on-sky demonstration by Sivo
of LQG for all modes in CANARY~\cite{Sivo:13}. Another
vibration scheme has been tested in GEMS by Guesalaga~\cite{Guesalaga:13}.
For use in GPI we have adapted the framework of Le Roux and 
added in the ability to work in a system with an arbitrary (i.e. non-integer frame) 
control delay~\cite{Poyneer2007Predictive-wave}, and also to correct both common-path 
and non-common-path vibrations~\cite{Poyneer2010Kalman-filterin},
both of which were deemed necessary to use an LQG controller in GPI.
As noted above, the tip-tilt and focus controllers have a 1.2 frame delay, which 
we account for explicitly in the LQG model.

%%-----------------------------------------------------------
%%-----------------------------------------------------------
\subsection{Modifications from original design} 
There are two significant differences between the tip-tilt LQG filter as initially
proposed and as used in GPI. First, in our proposal we presented a high-order
model for the atmospheric tip-tilt; in practice we always use a first-order auto-regressive (AR(1)) model.
Second, in GPI the pointing is actually controlled with two surfaces. By using a low-order
low-pass filter, the low temporal frequency part of the pointing signal is sent to the high-stroke
Stage, while the higher-frequency placed on the Woofer DM in combination with
the phase correction. Since each mirror has its own integrator, the result of the LQG
(which is integrated internally) must be converted to a pseudo-residual
that is then split and integrated on each surface. Specifically, if
the Woofer integrator has control law $C(\mathrm{z}) = 1/(1 - 0.999 \mathrm{z}^{-1})$, the
pseudo-residual is produced with a filter that is a``leaky'' differentiator: $D(z) = 1 - 0.999 \mathrm{z}^{-1}$.
We have had to be very conservative on the temporal split between the two surfaces
to avoid instabilities; the current cutoff for the split is at 25 Hz. 
A more robust solution would have been to have the  LQG model know explicitly about the 
temporal split~\cite{Correia2010Minimum-varianc}.

%%-----------------------------------------------------------
%%-----------------------------------------------------------
\subsection{On-sky results} 

\begin{figure}
   \begin{center}
   \begin{tabular}{c}
  \includegraphics[height=5cm]{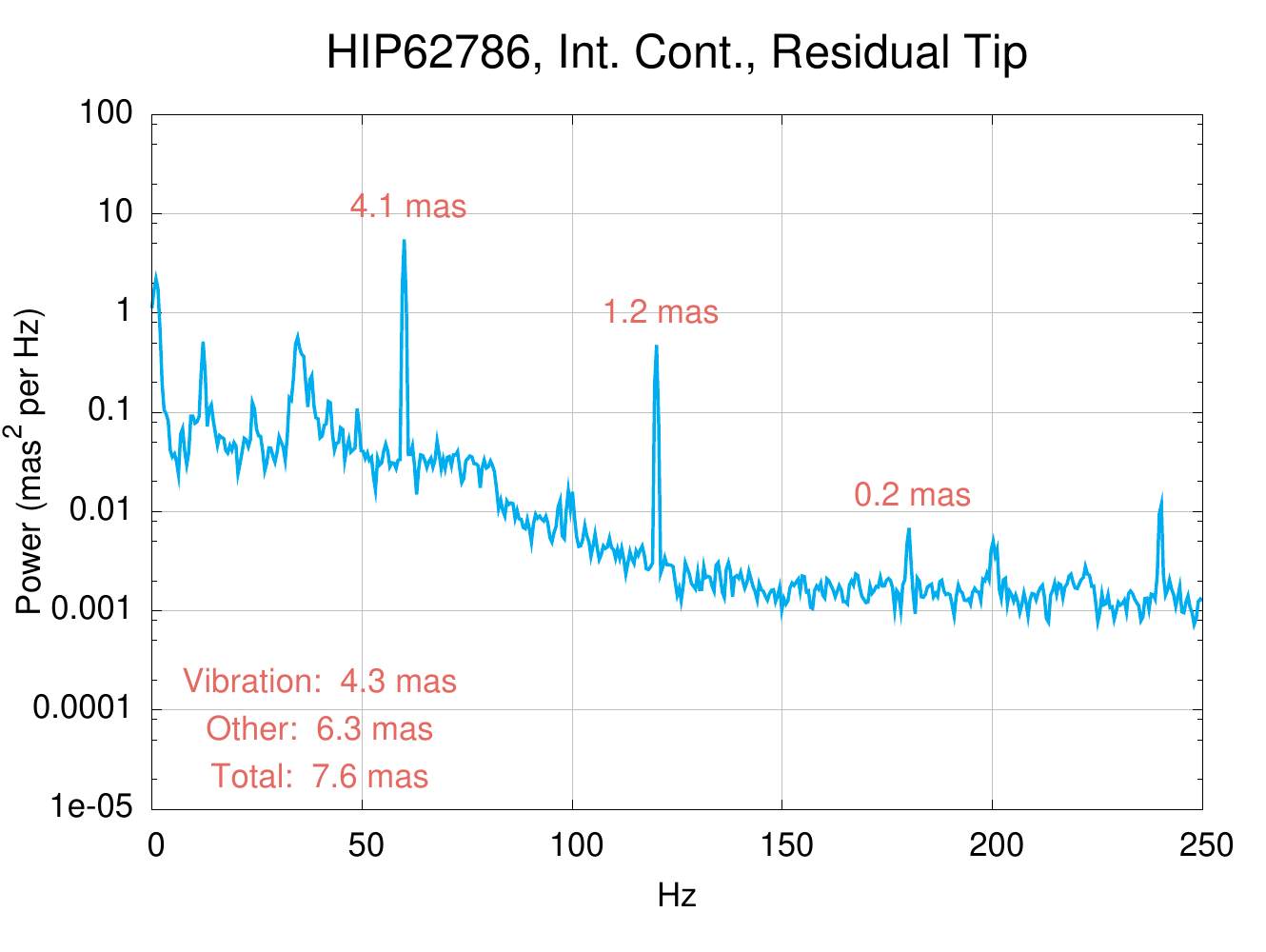}
  \includegraphics[height=5cm]{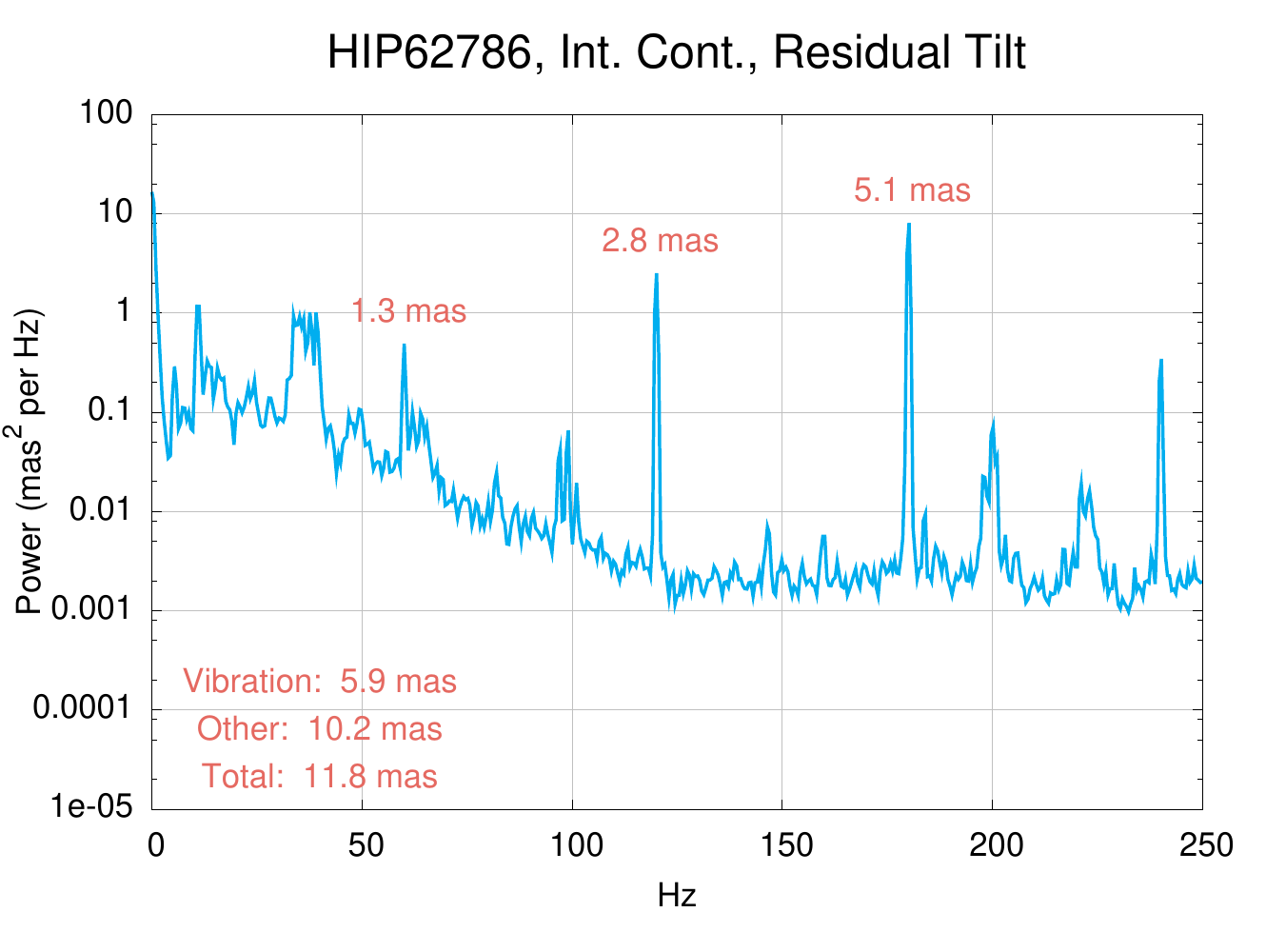}\\
  \includegraphics[height=5cm]{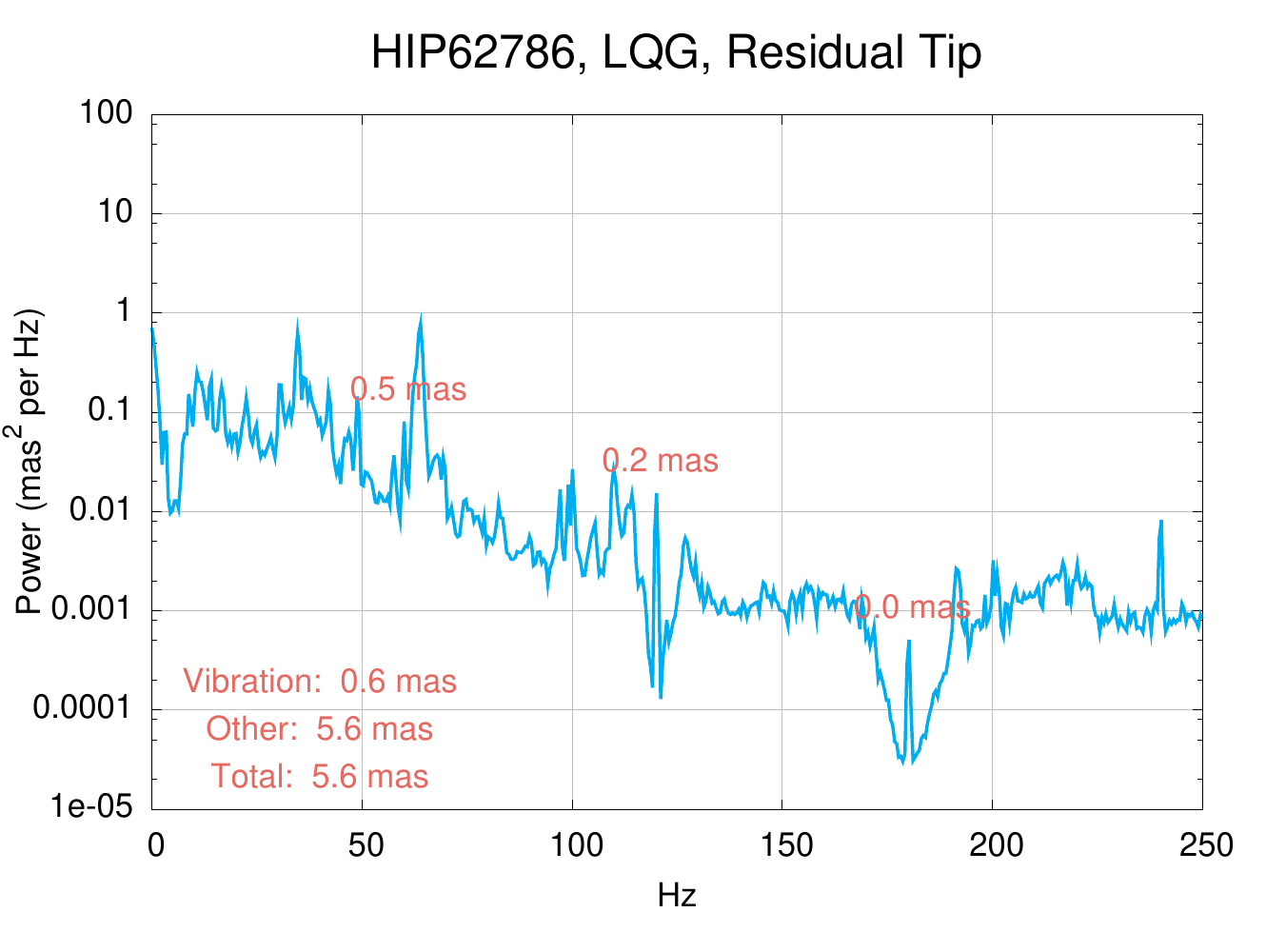}
  \includegraphics[height=5cm]{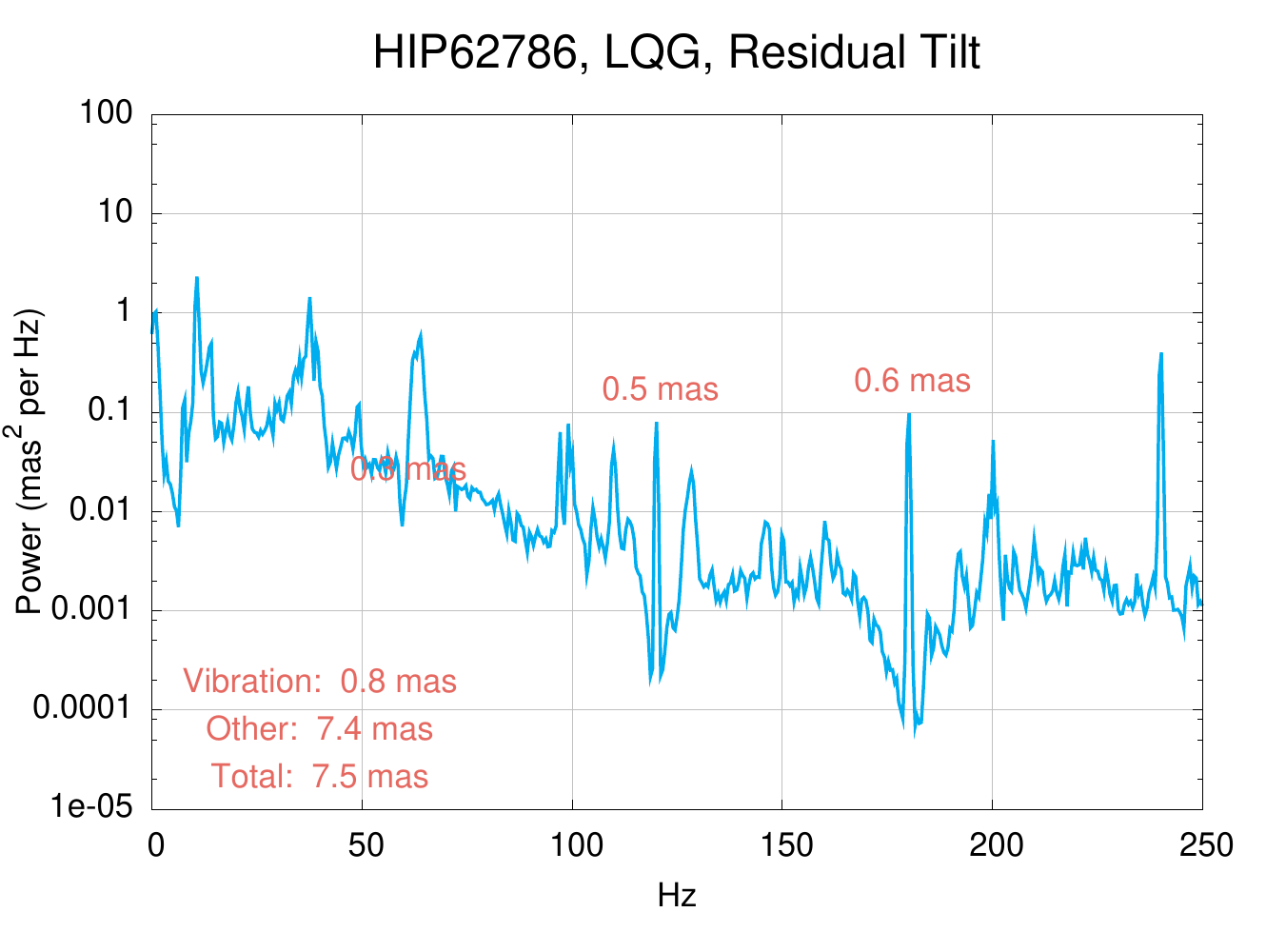}
   \end{tabular}
   \end{center}
   \caption[example] 
%>>>> use \label inside caption to get Fig. number with \ref{}
   { \label{fig:ttlqg} 
PSDs of closed-loop AO measurements for tip (left column) and tilt (right column), 
with estimated error amounts (red annotations). 
There are significant vibration-induced errors at 60, 120 and 180 Hz. 
With regular integral control (top row), these are 4 and 6 mas per axis. Use of a customized 
LQG controller (bottom row) reduces there vibration components to under 1 mas per axis.
The same LQG model was used for both axes. Telemetry: \_When\_2014.5.12\_19.50.22 and
\_When\_2014.5.12\_23.51.59.}
   \end{figure}

During integration and testing we identified the pointing vibrations at 60, 120 and 180 Hz as the
most significant. We were unable to determine if these tip-tilt vibrations are in the 
common path or the non-common path of the system, so we have assumed that 
they are common path and generated appropriate LQG filters.
After analyzing on-sky telemetry with the default integral controllers for pointing,
we generated LQG filters designed to recede the total vibration at 60, 120 and 180 Hz
to 1 mas.  LQG filters were tested in comparison to the 
default integral controller in an interleaved fashion.  Pointing vibrations are always
larger on the y-axis; in Figure~\ref{fig:ttlqg}
%-------------
%-------------
 we show the on-sky residual tip (top left)
and tilt (top right) as seen by 
the WFS for the integral controller. At bottom is the results for our preferred
LQG. The figure shows the AO measurements in closed loop. Annotated
in red are the error levels (see Section~\ref{sec:methods} for a discussion of measurements
versus error) passed on to science. The total vibration at 60, 120 and 180 Hz has
been reduced to less than 1 mas for each axis with use of the LQG.
The remaining pointing error is primarily residual atmosphere.

%%%%%%%%%%%%%%%%%%%%%%%%%%%%%%%%%%%%%%%%%%
%%%%%%%%%%%%%%%%%%%%%%%%%%%%%%%%%%%%%%%%%%
%%%%%%%%%%%%%%%%%%%%%%%%%%%%%%%%%%%%%%%%%%
\section{Focus vibration correction with LQG} \label{sec:foclqg}  % \label{} allows reference to this section

%%-----------------------------------------------------------
%%-----------------------------------------------------------
\subsection{Technology description and methods} 

When GPI is on the telescope, its cryocoolers cause vibration and 
result in significant phase errors. his manifests as a low-order wavefront - primarily focus, spherical and trefoil - 
with a temporal RMS of up to 150 nm occuring at 60 Hz. It remains unclear what is vibrating; 
either the secondary mirror moving in both piston and deforming, or a deformation of the 
primary mirror; both are very lightweight. 
For a thorough discussion of errors and mitigations,
see Hartung~\cite{hartungthis}. 
This phase vibration was a highly suitable candidate for LQG correction - 
the error was highly concentrated at 60 Hz, and analysis of the temporal variation
of the phase shape indicated that it was primarily focus that oscillated in total amplitude.

To correct this we simply applied the LQG filtering framework that had already been tested
for tip-tilt. In the tip-tilt case, the signal is calculated and removed from the centroids and 
then sent to the LQG and on to the Stage and Woofer. In the case of focus we calculate
focus directly from the slopes. To calculate and remove focus from the slopes via projection
we use a model to make x- and y-slope vectors for focus. To apply the focus correction
to the surface of the Woofer, we use the measured influence functions to generate a 
9 by 9 signal which makes focus. The slope and command signals are calibrated 
against each other to ensure that the overall loop gain is 1.

%%-----------------------------------------------------------
%%-----------------------------------------------------------
\subsection{On-sky results} 

\begin{figure}
   \begin{center}
   \begin{tabular}{c}
  \includegraphics[height=5cm]{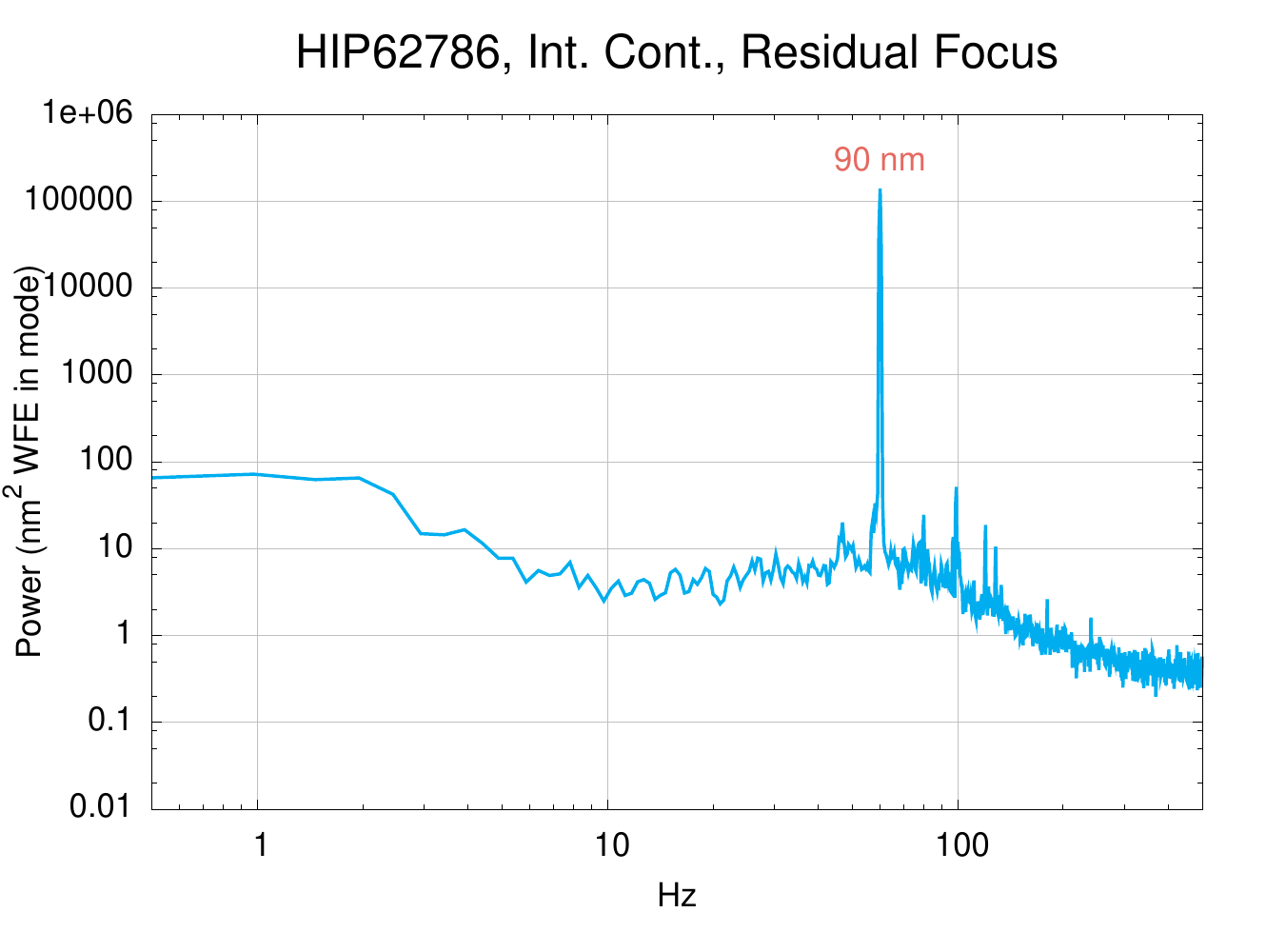}
  \includegraphics[height=5cm]{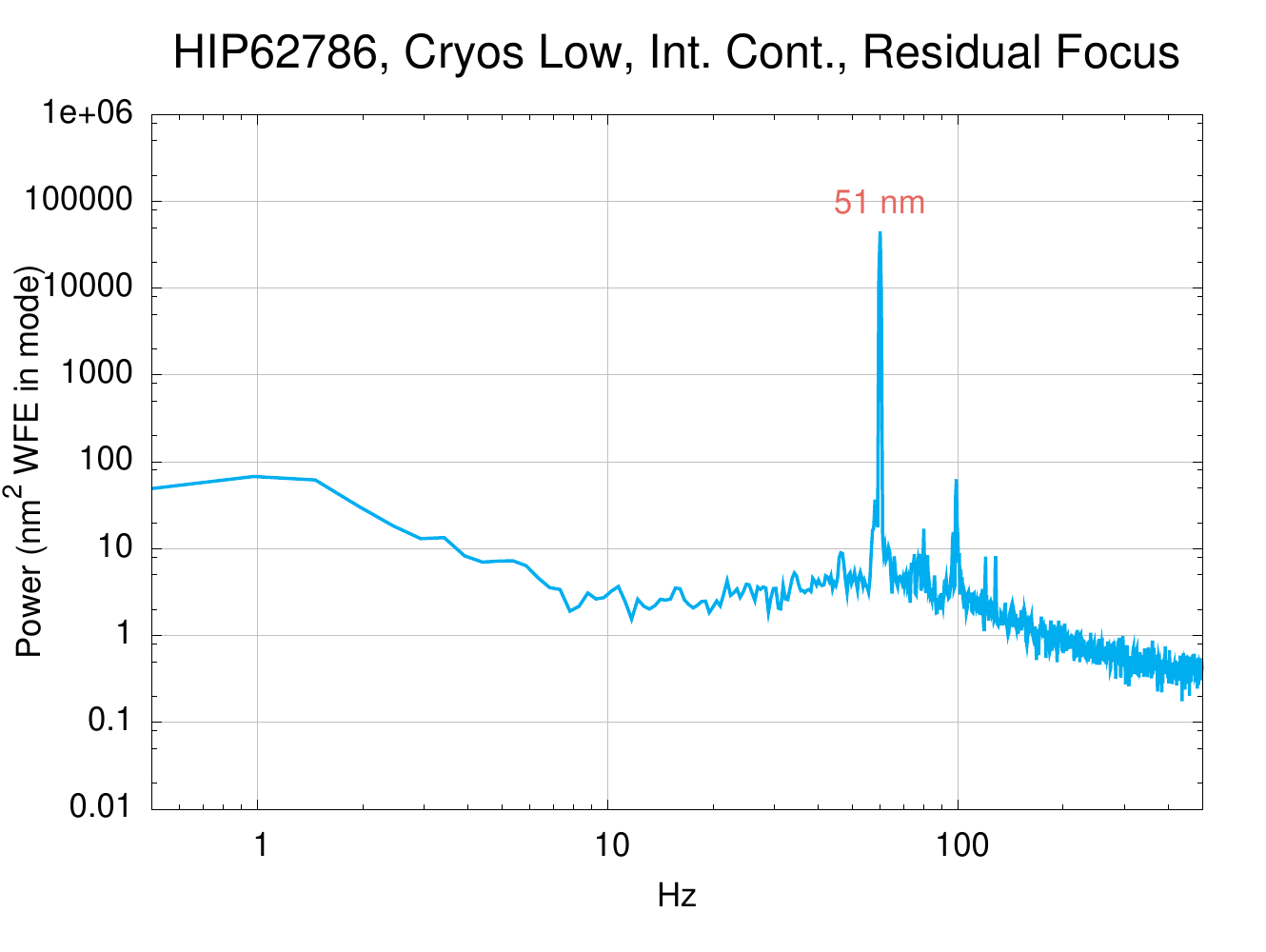}\\
  \includegraphics[height=5cm]{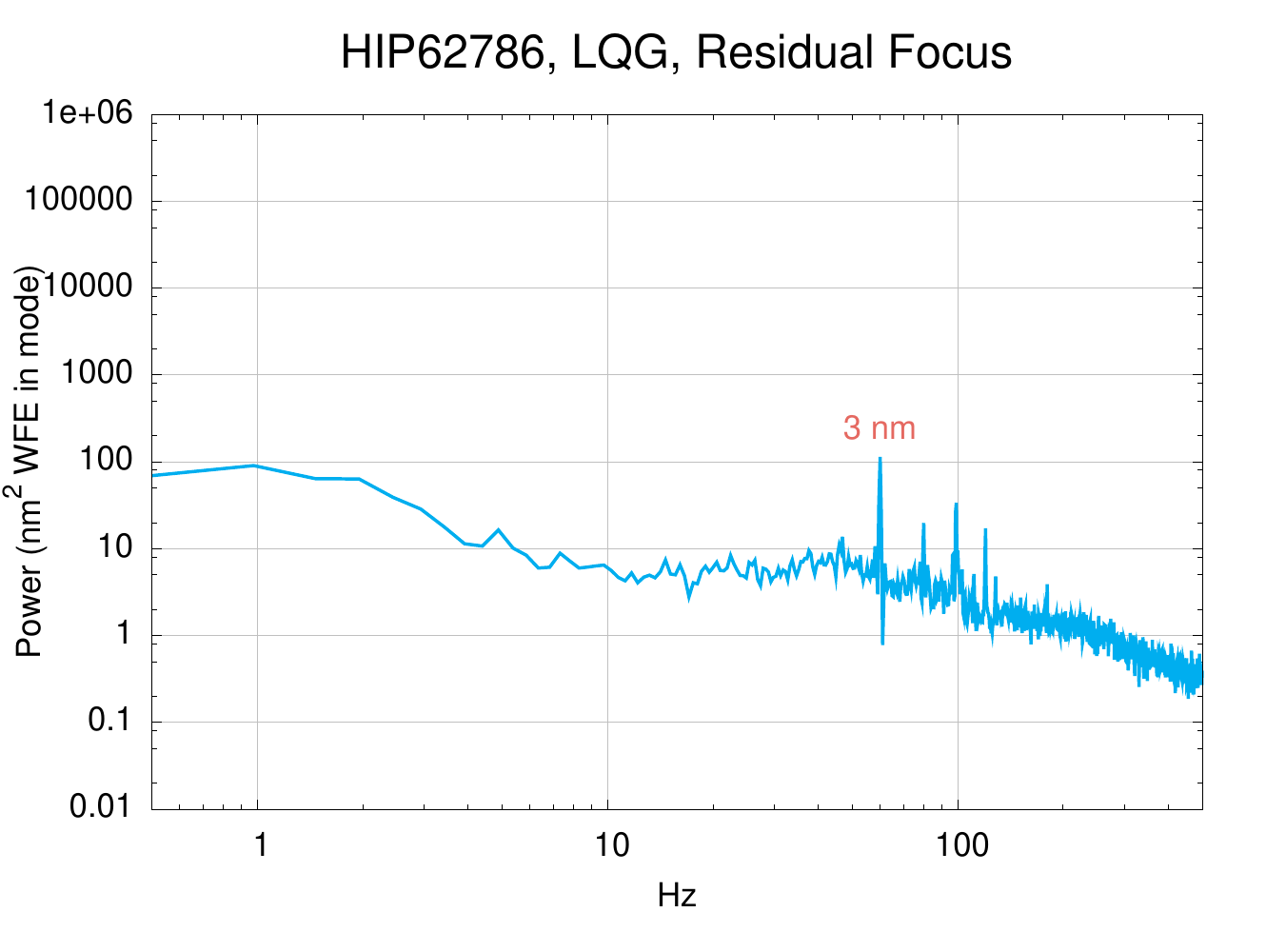}
   \end{tabular}
   \end{center}
   \caption[example] 
%>>>> use \label inside caption to get Fig. number with \ref{}
   { \label{fig:focuslqg} 
PSDs of closed-loop AO measurements for focus (blue), with estimated error amount
(red annotation) at 60 Hz. With cryocoolers on at full power (top left) and an integral controller, 
the 60 Hz focus error was 90 nm rms. With cryocoolers at low power (top right), the error is reduced to 51 nm rms.
Use of the LQG controller with cryocoolers at full power reduces the 60 Hz term to 3 nm rms, which is
now a negligible amount. Telemetry: \_When\_2014.5.13\_0.3.53,
    \_When\_2014.5.13\_0.25.28 and 
     \_When\_2014.5.13\_0.10.18.}
   \end{figure}

The focus LQG was tested on-sky during the May run. In this case a pure focus shape
was pulled out; the remaining components of the 60 Hz phase (mainly trefoil and spherical)
were sent through the high-order loop. Figure~\ref{fig:focuslqg} shows three
different temporal PSDs of the focus measurements. 
 %-------------
%-------------
Annotated in red is the estimate
error that is sent on to the science leg from the 60 Hz component. At upper left
is on-sky operation with an integral controller for the FOcus control. In this case the 
60 Hz error leads to a time-average of 90 nm rms wavefront error due to focus.
When the cryocoolers are turned down, but not off, the amount of focus at 60 Hz
is reduced to 51 nm. (When the cryocoolers are turned off completely this amount is
0 nm; due to system constraints we were not able to test this option as well on this target.)
With the cryocoolers at normal and the LQG, the deep notch (see Figure~\ref{fig:etf}, left panel)
strongly rejects the 60 Hz signal. In this case the result (bottom panel of figure) is
just 3 nm. This level is negligible in terms of the overall error budget.

Two more items remain in the validation of the focus control. First, we still do not
have definitive science images that show the difference between LQG on and off.
Second, the exact spatial shape of the error to be corrected, that is not just focus
but some combination of focus, trefoil and sphere, could be refined. 
 
%%%%%%%%%%%%%%%%%%%%%%%%%%%%%%%%%%%%%%%%%%
%%%%%%%%%%%%%%%%%%%%%%%%%%%%%%%%%%%%%%%%%%
%%%%%%%%%%%%%%%%%%%%%%%%%%%%%%%%%%%%%%%%%%
\section{Error budget} \label{sec:eb}  

We have developed a provisional error budget. These numbers are generated
from analysis of closed-loop AO measurements following the methods outlined
in Section~\ref{sec:methods} to convert from measurements to estimated error 
passed on to the science leg.   
 \begin{table}[h]
\caption{\label{tab:eb} Error budget. Important observational parameters
given in top panel. Error amounts in nm rms wavefront as determined from 
AO telemetry in second panel.
Telemetry used: a: \_When\_2014.3.22\_21.58.10;
b: \_When\_2013.12.11\_3.30.29;
c: \_When\_2014.3.22\_7.18.5;
d: \_When\_2014.3.24\_2.37.15;
e: \_When\_2014.5.14\_20.1.4.} 
\label{tab:fonts}
\begin{center}       
\begin{tabular}{|l|l|l|l|l|l|} %% this creates two columns
%% |l|l| to left justify each column entry
%% |c|c| to center each column entry
%% use of \rule[]{}{} below opens up each row
\hline
\rule[-1ex]{0pt}{3.5ex}  Target & $\beta$ Pictoris$^a$ & $\beta$ Pictoris$^b$  & HD141569$^c$ & HIP59563$^d$  
& HD101615$^e$ \\
\hline
\hline
\rule[-1ex]{0pt}{3.5ex}  Star I-band mag & 3.9 & 3.9 & 7 & 9 & 5.6 \\
\hline
\rule[-1ex]{0pt}{3.5ex}  Gains & OFC & OFC & 0.2 & OFC & OFC  \\
\hline
\rule[-1ex]{0pt}{3.5ex}  Frame rate & 1 kHz & 1 kHz & 1 kHz & 500 Hz & 1 kHz \\
\hline
\rule[-1ex]{0pt}{3.5ex}  cryocoolers &  On &  Low &  Off &  On  & On\\
\hline
\rule[-1ex]{0pt}{3.5ex}  LQG&  Off & Off &  Off & Off  & On\\
\hline
\rule[-1ex]{0pt}{3.5ex}  Seeing & best & normal & normal & normal & normal \\
\hline
\hline
\rule[-1ex]{0pt}{3.5ex}  Servo-lag error & 25 & 48 & 61 & 99 & 55   \\
\hline
\rule[-1ex]{0pt}{3.5ex}  AOWFS noise & 6 & 7 & 24 & 52  & 16  \\
\hline
\rule[-1ex]{0pt}{3.5ex}  Static error & 12 & 11 & 16 & 16 & 9   \\
\hline
\rule[-1ex]{0pt}{3.5ex}  60 Hz vibrations & 109 & 51 & 0 & 88 & 31   \\
\hline
%\rule[-1ex]{0pt}{3.5ex}  Total & X & X & X & X  & X  \\
%\hline
\end{tabular}
\end{center}
\end{table} 
Error estimates are given in Table~\ref{tab:eb}  for five different
observations. For each observation the star name and magnitude
is given. Modal gain optimization (``OFC'') was used in four of the five cases,
though on targets brighter than 7th this essentially is the
same as uniform gains of value 0.3. 

The first observation of $\beta$ Pictoris from March represents performance
in very good seeing - we were able to use the spatial filter at 2.8 mm
stably for several minutes, and servo lag error was estimated to be just 25 nm.
During more typical seeing on a bright to moderate target with loop gains at maximum,
the servo lag error is around 50 nm (second $\beta$ Pictoris, HD101615, HD141569). 

The HIP59563 observation (the same one used in Section~\ref{sec:ofc} for 
testing the gain optimizer) is representative of our performance on the 
dimmest  target required of GPI. At 500 Hz with optimized gains we have 99 nm
of servo lag error and 52 nm of WFS noise. 
With cryocoolers at normal power with no LQG, science leg error from the 60 Hz phase
is generally between 80 and 120 nm (depends on conditions and orientation). With
cryocoolers off this is reduced to 0; with the LQG as tested in May on the focus shape,
there are 31 nm of 60 Hz phase leftover - this is dominated by trefoil and spherical.
As noted above, this may be reduced further, but it is no longer a dominant term
in the AO error budget.

\begin{figure}
   \begin{center}
   \begin{tabular}{c}
  \includegraphics[height=5cm]{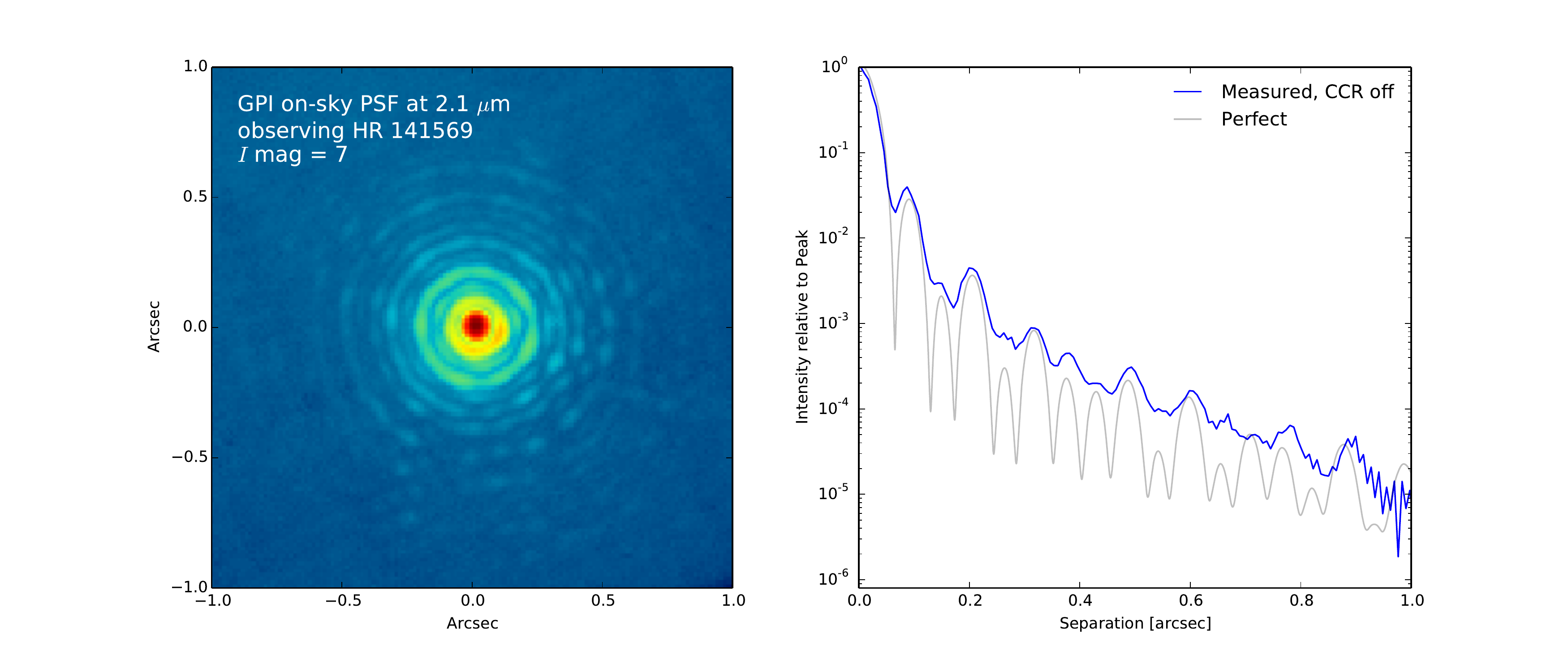}
   \end{tabular}
   \end{center}
   \caption[example] 
%>>>> use \label inside caption to get Fig. number with \ref{}
   { \label{fig:airypsf} 
GPI narrow-band point spread function at 2.1 microns for a 7th magnitude star, 
as observed with the GPI IFS~\cite{larkinthis} and reduced with the GPI Data 
Reduction Pipeline~\cite{perrinthis}.  Total exposure time was 45 s, obtained 
as 1.5 s * 10 coadds plus 5 s * 6 coadds for increased dynamic range. This is a 
synthetic narrow band image obtained by integrating over a small range of 
wavelengths in the data cube ($\lambda_c = 2.10 \mu$m, $\lambda/\Delta \lambda = 30$). 
In an azimuthally-averaged radial profile up to 15 Airy rings can be detected, 
as shown at right.  The grey curve shows the theoretical perfect PSF for the 
Gemini South telescope aperture geometry, integrated over the same spectral bandpass.}
   \end{figure}

The telemetry for HD141569 was taken co-incident with an non-coronagraphic image,
which is shown in Figure~\ref{fig:airypsf}.
 %-------------
%-------------
The Strehl of this image has not been rigorously estimated at this time. However, 
up to 15 Airy rings can be detected in the narrow-band image, indicating that the
AO system is performing well. We estimate the fitting~\cite{Hardy} and 
aliasing error\cite{RigautModel} terms from theory using the coefficient of 0.3, such
that $\sigma^2_\mathrm{fitting} = 0.3 (d/r_0)^{5/3}$ and 
$\sigma^2_\mathrm{aliasing} = 0.33\sigma^2_\mathrm{fitting}$.
We assume that $r_0$ was 14 cm for the typical seeing conditions.
For the HD141569 observation this gives us a fitting error of 54 nm and an aliasing error of
31 nm. Adding in the assumed 30 nm non-common-path errors given in Macintosh~\cite{Macintosh12052014},
we add in quadrature to the terms in the HD141569 column of the table:
 $(54^2 + 31^2 + 61^2  + 24^2 + 16^2 + 0^2 + 30^2)^{1/2} = 97 $ nm.
At 2.1 microns, this is a Strehl of 92\%.

We reiterate that we do not have a direct measurement of fitting error, aliasing error of 
non-common-path errors, but are using our best estimate at this time.
In future work we plan to improve it through
better estimates of seeing (and hence fitting), better examination of the aliasing
error through analysis of telemetry, and improved information on the non-common-path error.
These improved estimates will be accompanied by analysis of the co-incident science
images, which was beyond the scope of this work.

%%%%%%%%%%%%%%%%%%%%%%%%%%%%%%%%%%%%%%%%%%
%%%%%%%%%%%%%%%%%%%%%%%%%%%%%%%%%%%%%%%%%%
%%%%%%%%%%%%%%%%%%%%%%%%%%%%%%%%%%%%%%%%%%
\section{Conclusions} \label{sec:con}  
 GPI's AO system features several new technologies. During its verification and commissioning
 period we have systematically tested these technologies and analyzed performance. 
 The computationally efficient FTR method of wavefront reconstruction is successful, though
 we have to remove rotation from the centroids to prevent driving a scalloped shape on the 
 Tweeter when the spot size changes (due to either bias level or spot size due to seeing). 
 The modal gain optimization method updates 2304 modal gains for the Fourier modes
 every 8 seconds in closed loop. Tests have shown that the optimizer is correctly finding
 gains for each individual mode that minimize the error, and also minimizing overall error.
 Modal gain filters show clear evidence of wind. 

The AO system uses an LQG controller for tip, tilt and focus to reject vibrations. 
For tip and tilt we correct common-path vibrations at 60, 120 and 180 Hz to under 1 mas per axis.
By pulling focus out of the centroids and sending it through the LQG, we can very strongly
reject the focus at 60 Hz , reducing it to just 3 nm rms wavefront error. There is some 60 Hz phase
left over (trefoil and spherical) that could be further reduced if it helps overall performance.
The spatially-filtered wavefront
sensor has under-performed relative to our expectations. 
We have clear evidence, thanks to layers of frozen flow turbulence, that the 
spatial filter does reject aliases. However, we are not able to regularly use the spatial filter
stably at it nominal $\lambda/d$ diameter. In very good seeing we can get it down to 2.8 mm,
but it is more typical that we use it at 3.5 mm. At this size it clears out aliasing from the 
inner 42\% of the dark hole (i.e. to radius $0.42\lambda/d$ instead of $\lambda/d$.)
We have presented an provisional error budget of AO terms that we can directly measure.
On bright to moderate targets (down to around 7th magnitude), performance is
dominated by servo lag error. In typical seeing this term is 50 nm. On dimmer targets
the gain optimizer will trade off servo lag and WFS noise. On our dimmest required target
of 9th magnitude we achieved 99 nm servo lag error and 52 nm WFS noise.

We will continue to work during the end of the verification and commissioning period to
analyze existing science images and obtain new ones as necessary to  confirm
our analysis, which is here based on AO telemetry. Of particular interest are 
verifying the impact of the focus LQG, which on most targets removes the dominant
60 Hz term in the error budget. We also want to quantify contrast in longer exposures
as a function of spatial filter size and determine if we can do better than 3.5 mm, which is
oversized by about 60\%.

%%%%%%%%%%%%%%%%%%%%%%%%%%%%%%%%%%%%%%%%%%%%%%%%%%%%
%\appendix    %>>>> this command starts appendixes
%%%%%%%%%%%%%%%%%%%%%%%%%%%%%%%%%%%%%%%%%%%%%%%%%%%%%
%\section{Optional Appendix} \label{sec:misc}
%
%Blah blah.

%%%%%%%%%%%%%%%%%%%%%%%%%%%%%%%%%%%%%%%%%%%%%%%%%%%%%%%%%%%%%
\acknowledgments     %>>>> equivalent to \section*{ACKNOWLEDGMENTS}       

This work performed under the auspices of the U.S. Department of Energy 
by Lawrence Livermore National Laboratory under Contract DE-AC52-07NA27344.
The document number is LLNL-CONF-655984.
The Gemini Observatory is operated by the 
Association of Universities for Research in Astronomy, Inc., under a cooperative agreement 
with the NSF on behalf of the Gemini partnership: the National Science Foundation 
(United States), the National Research Council (Canada), CONICYT (Chile), the Australian 
Research Council (Australia), Minist\'{e}rio da Ci\^{e}ncia, Tecnologia e Inova\c{c}\~{a}o 
(Brazil) and Ministerio de Ciencia, Tecnolog\'{i}a e Innovaci\'{o}n Productiva (Argentina).

%%%%%%%%%%%%%%%%%%%%%%%%%%%%%%%%%%%%%%%%%%%%%%%%%%%%%%%%%%%%%
%%%%% References %%%%%

%\bibliography{refs}   %>>>> bibliography data in report.bib
%\bibliographystyle{spiebib}   %>>>> makes bibtex use spiebib.bst

\end{document}